\documentclass[12pt]{article}

\newcommand{\resection}[1]{\setcounter{equation}{0}\section{#1}}

\begin{document}
\setcounter{page}{0} \topmargin 0pt \oddsidemargin 5mm
\renewcommand{\thefootnote}{\arabic{footnote}}
\newpage
\setcounter{page}{0}
\begin{titlepage}

\begin{flushright}
SISSA/16/2001/FM
\end{flushright}

\vspace{0.5cm}
\begin{center}
{\large {\bf Boundary bootstrap principle in two-dimensional integrable quantum field theories}} \\
\vspace{2cm}
{\large Valentina Riva$^{1,2}$} \\
\vspace{0.5cm} {\em $^{1}$Dipartimento di Scienze Chimiche, Fisiche e Matematiche,
Universit\`a dell'Insubria, Como, Italy}\\
\vspace{0.3cm} {\em $^{2}$International School for Advanced Studies, Trieste, Italy }
\end{center}
\vspace{1cm}

\renewcommand{\thefootnote}{\arabic{footnote}}
\setcounter{footnote}{0}

\begin{abstract}

We study the reflection amplitudes of affine Toda field theories with boundary, following the ideas developed by
Fring and Koberle in \cite{fring0},\cite{fring1},\cite{fring2} and focusing our attention on the $E_{n}$ series
elements, because of their interesting structure of higher order poles.

We also investigate the corresponding minimal reflection matrices, finding, with respect to the bulk case, a more
complicated relation between the spectra of bound states associated to the minimal and to the \lq\lq
dressed\rq\rq \, amplitudes.

\end{abstract}

\vspace{1cm}

\end{titlepage}

\newpage

\resection{Introduction}

A two-dimensional quantum field theory with boundary can be basically defined in two ways (\cite{ghoszam}). Let us
consider the semi-infinite euclidean plane, $x\in (-\infty,0]$, $y\in (-\infty,\infty)$, where the $y$-axis
represents the boundary. In the Lagrangian approach one writes the action in the form
\begin{equation}\label{lagr}
{\cal A}=\int_{-\infty}^{\infty}dy\int_{-\infty}^{0}dx\:
a\left(\varphi,\partial_{\mu}\varphi\right)+\int_{-\infty}^{\infty}dy
\:b\left(\varphi_{B},\frac{d}{dy}\varphi_{B}\right),
\end{equation}
where $a$ and $b$ are local functions respectively of the bulk and boundary \lq\lq fundamental fields\rq\rq \,
$\varphi,\varphi_{B}$ ($\varphi_{B}(y)=\varphi(x,y)|_{x=0}$). In the \lq\lq perturbed conformal field theory\rq\rq
\, approach one writes the symbolic action
\begin{equation}\label{pert}
{\cal A}={\cal A}_{CFT+CBC}+\int_{-\infty}^{\infty}dy\int_{-\infty}^{0}dx\: \Phi(x,y)+\int_{-\infty}^{\infty}dy
\:\Phi_{B}(y),
\end{equation}
where ${\cal A}_{CFT+CBC}$ is the action of a conformal field theory (CFT) on the semi-infinite plane with certain
conformal boundary conditions (CBC), and $\Phi,\Phi_{B}$ are specific bulk and boundary relevant fields.

If a \lq\lq bulk\rq\rq \, theory is integrable, i.e. if it possesses an infinite set of mutually commutative
integrals of motion, the scattering processes are purely elastic, and the $S$-matrix factorizes into two-particle
amplitudes. This can be generalized to the case with boundary defining, in addition to the bulk $S$-matrix, an
opportune boundary reflection matrix which will factorize into single-particle amplitudes. In every integrable
boundary field theory we expect to find a certain number of different reflection matrices, corresponding to the
various boundary conditions compatible with integrability.

In the \lq\lq perturbed conformal field theory\rq\rq \, approach the integrable boundary conditions must be
associated in the ultraviolet limit to some of the possible conformal ones. It was demonstrated by Cardy
(\cite{cardy}) that, in the case of Virasoro minimal models, conformal boundary conditions are in one-to-one
correspondence with the primary fields of the examined CFT. In fact, imposing the absence of energy or momentum
flux across the boundary, one can write the physical boundary states as
\begin{equation}\label{confstates}
|\tilde{l}\rangle=\sum_{j}\frac{S_{lj}}{\sqrt{S_{0j}}}|j\rangle\rangle,
\end{equation}
where the indices $l$ and $j$ label the primary operators of the theory ($0$ refers to the identity), $S_{lj}$ is
the modular matrix and the states $|j\rangle\rangle$, called \lq\lq Ishibashi states\rq\rq , are constructed from
the states in the irreducible quotient of the Verma module $j$. A fundamental quantity related to a boundary state
$|A\rangle$ is the so-called ground state degeneracy $g$, defined as (\cite{affleck})
\begin{equation}\label{g}
g_{A}=\langle 0|A\rangle,
\end{equation}
where $|0\rangle$ is the ground state of the bulk Hamiltonian. If the state $|A\rangle$ is of the form
(\ref{confstates}), $g$ has the generally noninteger value
\begin{equation}\label{valg}
g_{A}=\frac{S_{A0}}{\sqrt{S_{00}}}.
\end{equation}

\vspace{0.5cm}

In the bulk case, the two-particle scattering amplitudes can be exactly calculated (up to the so-called \lq\lq CDD
ambiguity\rq\rq) as solutions of a number of constraints, which are expressed by unitarity, crossing, Yang-Baxter
and bootstrap equations. In the presence of a boundary, the reflection amplitudes are constrained by analogous
equations, which relate them to the bulk $S$-matrix. In this way, starting from a theory whose bulk scattering
matrix is known, one can investigate which are the possible reflection amplitudes.

If the examined theory is a perturbed minimal model, we know explicitly from the primary fields content of the
corresponding CFT which are the conformal boundary conditions that we have to recover in the UV limit.

On the contrary, the affine Toda field theory related to an algebra ${\cal G}$ of rank $r$ has in general a
description in terms of the Lagrangian
\begin{equation}\label{toda}
{\cal
L}=\frac{1}{2}\sum_{j=1}^{r}\left(\partial_{\mu}\phi^{j}\right)^{2}-\frac{m^{2}}{\beta^{2}}\sum_{i=0}^{r}n_{i}e^{\beta\alpha_{i}\cdot\phi},
\end{equation}
where the $\phi^{j}$'s are $r$ bosonic fields, $m$ and $\beta$ are real parameters, $\alpha_{i}$ are the simple
roots of ${\cal G}$, and $\alpha_{0}=-\sum_{i=1}^{r}n_{i}\alpha_{i}$ is the opposite of the highest root. We can
now define a boundary field theory with Lagrangian
\begin{equation}\label{todabound}
{\cal L}_{B}=\theta(-x){\cal L}-\delta(x){\cal B}(\phi)
\end{equation}
and investigate which reflection amplitudes are compatible with the bulk $S$-matrix. Restrictions on the boundary
interaction ${\cal B}(\phi)$ due to the integrability requirement have been discussed in \cite{dur},\cite{dur2}.
Also this theory can be seen as a perturbation of a certain CFT, which has however a spectrum of primary operators
generally much richer than in minimal models. Furthermore, this CFT possesses a larger symmetry than just
conformal invariance, and in this case the structure of the boundary states for a boundary condition that is only
conformally invariant is not yet known.

\newpage

\resection{Boundary Affine Toda Field Theories}

In a two-dimensional integrable field theory it is possible to parameterize the momentum of a particle in terms
of a single variable $\theta$, called \textit{rapidity}:
\begin{equation}\label{rap}
p_{a}=m_{a}\left(\cosh\theta_{a},\sinh\theta_{a}\right).
\end{equation}
Assuming that the effect of the boundary is to reverse the momentum and preserve the energy of a particle which
scatters off it, we define the reflection amplitude $K(\theta)$ as the proportionality factor which relates a
single-particle \lq out\rq \, state to the \lq in\rq \, state with reversed momentum:
\begin{equation}
|a,\theta\rangle_{out}=K_{a}\left(\theta\right)|a,-\theta\rangle_{in}.
\end{equation}
With this definition we restrict our attention to theories with distinguishable particles, otherwise the
reflection factor $K_{a}$ should be a matrix to allow for a mixing of states.

The main equations for the reflection amplitudes have been derived in \cite{ghoszam} and discussed in
\cite{fring0}-\cite{sas},\cite{corr}. The unitarity condition
\begin{equation}\label{unit}
K_{a}\left(\theta\right)K_{a}\left(-\theta\right)=1
\end{equation}
is a straightforward generalization of the bulk one, while the boundary crossing equation
\begin{equation}\label{cross}
K_{a}\left(\theta\right)K_{\bar{a}}\left(\theta+i\pi\right)=S_{aa}\left(2\theta\right)
\end{equation}
required more attention, and will be of great importance in our considerations.

\vspace{3cm}

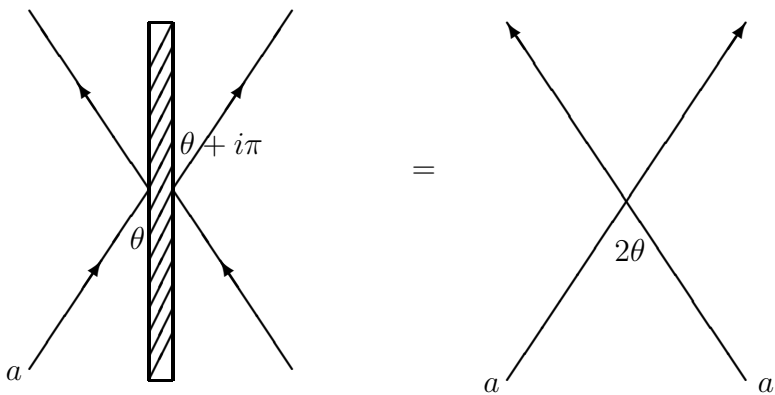
\begin{figure}[h]
\setlength{\unitlength}{0.0125in}
\begin{picture}(40,90)(60,420)

\thicklines \put(200,420){\line( 0,1){150}} \put(210,420){\line( 0,1){150}} \put(200,420){\line( 1,0){10}}
\put(200,570){\line( 1,0){10}}

\put(200,420){\line( 1,2){10}} \put(200,430){\line( 1,2){10}} \put(200,440){\line( 1,2){10}} \put(200,450){\line(
1,2){10}} \put(200,460){\line( 1,2){10}} \put(200,470){\line( 1,2){10}} \put(200,480){\line( 1,2){10}}
\put(200,490){\line( 1,2){10}} \put(200,500){\line( 1,2){10}} \put(200,510){\line( 1,2){10}} \put(200,520){\line(
1,2){10}} \put(200,530){\line( 1,2){10}} \put(200,540){\line( 1,2){10}} \put(200,550){\line( 1,2){10}}

\put(200,500){\line(-2,-3){20}} \put(150,425){\vector(2,3){30}}

\put(140,420){$a$} \put(192,475){$ \theta$}

\put(200,500){\vector(-2,3){30}} \put(170,545){\line(-2,3){20}}

\put(210,500){\line(2,-3){20}} \put(260,425){\vector(-2,3){30}}

\put(210,500){\vector(2,3){30}} \put(240,545){\line(2,3){20}}

\put(450,420){\vector(-2,3){100}} \put(350,420){\vector(2,3){100}}

\put(395,470){$ 2\theta$} \put(213,515){$ \theta + i \pi$} \put(455,415){$a$} \put(340,415){$a$}
\put(310,505){$=$}
\end{picture}
\caption{The boundary crossing-unitarity relation}
 \end{figure}

\vspace{0.5cm}

In the case of distinguishable particles the Yang-Baxter equations
\begin{equation}\label{YB}
K_{a}\left(\theta_{a}\right)S_{ab}\left(\theta_{b}+\theta_{a}\right)K_{b}\left(\theta_{b}\right)S_{ab}\left(\theta_{b}-\theta_{a}\right)=
S_{ab}\left(\theta_{b}-\theta_{a}\right)K_{b}\left(\theta_{b}\right)S_{ab}\left(\theta_{b}+\theta_{a}\right)K_{a}\left(\theta_{a}\right)
\end{equation}
are trivially satisfied. On the contrary, a strong constraint is given by the bootstrap equations
\begin{equation}\label{boot}
K_{c}\left(\theta\right)= K_{a}(\theta-i \bar{u}_{ac}^{b})S_{ab}(2\theta-i \bar{u}_{ac}^{b}+i
\bar{u}_{bc}^{a})K_{b}(\theta+i \bar{u}_{bc}^{a})\,,
\end{equation}
with $\bar{u}_{ab}^{c}\equiv\pi-u_{ab}^{c}$, where the bulk \lq\lq fusing angles\rq\rq \, $u_{ab}^{c}$ correspond
to poles of $S_{ab}$ at $\theta=i u_{ab}^{c}$ creating the bound state $c$, and are related to the masses of the
particles by
\begin{equation}\label{carnot}
m_{c}^{2}=m_{a}^{2}+m_{b}^{2}+2m_{a}m_{b}\cos u_{ab}^{c}\,.
\end{equation}

\vspace{0.5cm}

In affine Toda field theories the $S$-matrix elements are always of the form $\prod_{x}\{x\}_{\theta}$, with
\begin{equation}\label{def}
\{x\}_{\theta}=\frac{s_{\frac{x+1}{h}}(\theta)s_{\frac{x-1}{h}}(\theta)}{s_{\frac{x+1-B}{h}}(\theta)s_{\frac{x-1+B}{h}}(\theta)}
, \qquad s_{\alpha}(\theta)=\frac{\sinh \left[\frac{1}{2}\left(\theta+i\pi \alpha\right)\right]}{\sinh
\left[\frac{1}{2}\left(\theta-i\pi\alpha\right)\right]}
\end{equation}
($B(\beta)\in[0,2]$ is a coupling constant dependent parameter, and $h$ is the Coxeter number of the algebra
${\cal G}$). Fring and Koberle (\cite{fring1}) have demonstrated that in this case the boundary crossing equation
(\ref{cross}) has factorized solutions of the form
\begin{equation}\label{K}
K_{a}(\theta)=\prod_{x}{\cal K}_{x}(\theta),\qquad {\cal
K}_{x}(\theta)=\frac{s_{\frac{1-x-h}{2h}}(\theta)s_{\frac{-1-x-h}{2h}}(\theta)}{s_{\frac{1-x-B-h}{2h}}(\theta)s_{\frac{-1-x+B-h}{2h}}(\theta)}
\end{equation}
where the blocks ${\cal K}_{x}(\theta)$ are in one-to-one correspondence to the blocks $\{x\}_{2\theta}$ in
$S_{a\bar{a}}$ up to a shift of $2h$ in $x$. In order to determine which of the blocks are shifted by $2h$ and
which are not, one has to write a bootstrap equation (\ref{boot}) involving only one particular $K_{a}(\theta)$,
find its most general solution, and then compute the other $K_{b}(\theta)$ consistently.

\vspace{0.5cm} Given a solution $K_{a}(\theta)$ of the equations (\ref{unit})-(\ref{boot}), a function of the form
\begin{equation}
K_{a}(\theta)\psi_{a}(\theta)
\end{equation}
is again a solution if $\psi_{a}$ satisfies the homogeneous equations
\begin{equation}\label{CDD1}
\psi_{\bar{a}}(\theta+i\frac{\pi}{2})\psi_{a}(\theta-i\frac{\pi}{2})=1,
\end{equation}
\begin{equation}\label{CDD2}
\psi_{c}(\theta)=\psi_{a}(\theta-i\bar{u}_{ac}^{b})\psi_{b}(\theta+i\bar{u}_{bc}^{a}).
\end{equation}
Some possible choices of these functions (called CDD-factors and analyzed in \cite{sas}) are

\begin{enumerate}

\item {$\psi_{a}(\theta)=\frac{K_{a}(\theta+i\pi)}{K_{a}(\theta)}$}

Every block $\psi_{x}(\theta)=\frac{{\cal K}_{x}(\theta+i\pi)}{{\cal K}_{x}(\theta)}=\frac{{\cal
K}_{x+2h}(\theta)}{{\cal K}_{x}(\theta)}$ satisfies eq. (\ref{CDD1}) and (\ref{CDD2}), so that the new solution
can be obtained from the first one by shifting all the $x$'s by $2h$.

\item {$\psi_{a}(\theta)=\left(S_{ab}(\theta)\right)^{\pm 1}$, $\forall b$  }

\item {$\psi_{a}(\theta)=\prod_{b}S_{ab}(\theta)$ }

\end{enumerate}
What we expect is that these CDD-factors relate different sets of reflection amplitudes corresponding to the
various boundary conditions compatible with integrability.

For the theories in exam we will analyze the two solutions of the form (\ref{K}) $K_{a}(\theta)$ and
$K_{a}(\theta+i\pi)$, related by the first kind of CDD-ambiguity mentioned, calling \lq\lq minimal\rq\rq \, the
one with the minimum number of poles in the physical strip.

\vspace{0.5cm}

As noticed in \cite{ghoszam}, the boundary can exist in several stable states, and the presence of poles in
$K_{a}(\theta)$ indicates the possibility for particle $a$ to excite it. For reflection amplitudes, the physical
strip is the region $0\leq \rm{Im}\,\theta\leq\frac{\pi}{2}$ of the complex $\theta$-plane.

\vspace{5cm}

\begin{figure}[h]
\setlength{\unitlength}{0.0125in}
\begin{picture}(40,0)(60,470)
\thicklines \put(330,460){\line( 0,1){160}}

\put(330,460){\line(2,1){20}}

\put(330,480){\line(2,1){20}} \put(330,490){\line(2,1){20}} \put(330,500){\line(2,1){20}}
\put(330,510){\line(2,1){20}} \put(330,520){\line(2,1){20}}

\put(330,550){\line(2,1){20}} \put(330,560){\line(2,1){20}} \put(330,570){\line(2,1){20}}
\put(330,580){\line(2,1){20}} \put(330,590){\line(2,1){20}}

\put(330,610){\line(2,1){20}}

\put(338,473){$\alpha $} \put(338,535){$\beta $} \put(338,603){$\alpha $}
\put(330,510){\line(-2,-3){10}} \put(304,471){\vector(2,3){16}} \put(295,605){$a$}

\put(313,478){$\eta_{a \alpha}^{\beta}$}

\put(330,570){\vector(-2,3){16}} \put(314,594){\line(-2,3){10}} \put(295,470){$a$}

\end{picture}
 \caption{Boundary excitation}
 \end{figure}
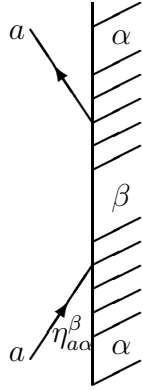

\vspace{0.5cm}

We will call $\eta_{a\alpha}^{\beta}$ the \lq\lq fusing angle\rq\rq \, related to an odd-order pole with positive
residue at $\theta_{a}=i\eta_{a\alpha}^{\beta}$ in the reflection amplitude $K_{a}^{\alpha}$ of particle $a$ on
the boundary state $\alpha$, creating the state $\beta$. The corresponding bootstrap equation for the boundary
bound states is
\begin{equation}\label{boundst}
K_{b}^{\beta}\left(\theta\right)=S_{ab}\left(\theta+i\eta_{a\alpha}^{\beta}\right)K_{b}^{\alpha}\left(\theta\right)S_{ab}\left(\theta-i\eta_{a\alpha}^{\beta}\right),
\end{equation}
and the energies of the two states $|\alpha\rangle$ and $|\beta\rangle$ are related by
\begin{equation}\label{energy}
E_{\beta}=E_{\alpha}+m_{a}\cos\left(\eta_{a\alpha}^{\beta}\right).
\end{equation}
The blocks ${\cal K}_{x}(\theta)$ have poles at
\begin{equation}\label{poles}
\theta_{\pm}=\frac{\pm 1-x-h}{2h}i\pi \;(\textrm{mod} \, 2\pi i)\qquad \textrm{and} \qquad
\theta_{\pm}^{B}=\frac{\pm B\mp 1 +x+h}{2h}i\pi \;(\textrm{mod} \, 2\pi i)
\end{equation}
and zeroes at
\begin{equation}\label{zeroes}
^{0}\theta_{\pm}=\frac{\pm 1+x+h}{2h}i\pi \;(\textrm{mod} \, 2\pi i)\qquad \textrm{and} \qquad
^{0}\theta_{\pm}^{B}=\frac{\pm 1\mp B -x-h}{2h}i\pi \;(\textrm{mod} \, 2\pi i).
\end{equation}
Only shifted blocks can give $\theta_{\pm}$ poles and $^{0}\theta_{\pm}^{B}$ zeroes in the physical strip.
Contrary to the case of the bulk $S$-matrix, the coupling constant dependent $\theta_{\pm}^{B}$ poles of the
unshifted blocks can move inside the strip $0\leq \rm{Im}\,\theta\leq\pi$, but they are confined inside the
interval $\frac{\pi}{2}\leq \rm{Im}\,\theta \leq\pi$ (they can reach the value $i\frac{\pi}{2}$ only in the
trivial cases $B=0,2$); the $^{0}\theta_{\pm}$ zeroes of unshifted blocks are also located in $\frac{\pi}{2}\leq
\rm{Im}\,\theta \leq\pi$.

\vspace{0.5cm}

Fring and Koberle have demonstrated in \cite{fring2} that every excited state reflection amplitude obtained by
eq.(\ref{boundst}) can be expressed in terms of the ground state one as
\begin{equation}\label{colour}
K_{a}^{\mu}\left(\theta\right)=\left(\prod_{b}S_{ab}\left(\theta\right)\right)K_{a}^{0}\left(\theta\right),
\end{equation}
where all the $b$'s are of the same colour with respect to the bicolouration of the Dynkin diagram of ${\cal G}$,
and the corresponding energy levels are related by
\begin{equation}\label{level}
E_{\mu}=E_{0}+\frac{1}{2}\sum_{b}m_{b}.
\end{equation}

\vspace{0.5cm}

\subsection{$E_{6}$-Affine Toda Field Theory}

A complete analysis of the \lq\lq minimal\rq\rq \, solution was performed in \cite{fring2}. The bootstrap closes
on eight boundary bound states, and there are six different energy levels, two of which are degenerate.

The bulk S-matrix of this model has been found in \cite{E6}. We label the six particles according to the following
Dynkin diagram

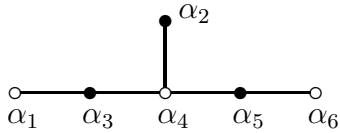
\setlength{\unitlength}{0.01cm}
\begin{picture}( 1000,300)(0,100)
\thicklines \put(492,190){$ \circ$} \put(510,200){\line( 1, 0){85}} \put(592,190){$ \bullet$}
\put(702,207){\line( 0, 1){85}} \put(692,285){$ \bullet$} \put(610,200){\line( 1, 0){85}} \put(692,190){$ \circ$}
\put(792,190){$ \bullet$} \put(892,190){$ \circ$} \put(710,200){\line( 1, 0){85}} \put(810,200){\line( 1, 0){85}}
\put(492,160){$ \alpha_1$} \put(592,160){$ \alpha_3$} \put(692,160){$ \alpha_4$} \put(792,160){$ \alpha_5$}
\put(892,160){$ \alpha_6$} \put(720,300){$ \alpha_2$}
\end{picture}

whose extension by $\alpha_{0}$ possesses a $Z_{3}$ symmetry, being invariant under the transformations
\begin{equation}
\alpha_{1}\rightarrow\alpha_{6}\rightarrow\alpha_{0},\quad
\alpha_{2}\rightarrow\alpha_{3}\rightarrow\alpha_{5},\quad \alpha_{4}\rightarrow\alpha_{4}.
\end{equation}

Omitting the $\theta$-dependence, the \lq\lq minimal\rq\rq \, set of reflection amplitudes is:

\begin{eqnarray}
\label{min1}K_{1}&=& {\cal K}_{5}{\cal K}_{35}  \\
\label{min2}K_{2}&=& {\cal K}_{1}{\cal K}_{7}{\cal K}_{11} {\cal K}_{29} \\
\label{min3}K_{3}&=& {\cal K}_{3}{\cal K}_{5}{\cal K}_{7}{\cal K}_{11}{\cal K}_{29}{\cal K}_{33} \\
\label{min4}K_{4}&=& {\cal K}_{1}{\cal K}_{3}{\cal K}^{2}_{5}{\cal K}_{7}{\cal K}^{2}_{9}{\cal K}_{27}{\cal
K}_{29} {\cal K}^{2}_{31}{\cal K}_{35}
\end{eqnarray}
with $K_{6}=K_{1}$ and $K_{5}=K_{3}$.

Fixing $E_{0}=0$, the energy levels are:
\begin{center}
\begin{tabular}{cclc|}
$E_{\alpha}$ &=& $0.796 m =\frac{m_{2}}{2}$ \\
$E_{\beta}$ &=& $1.087 m =\frac{m_{3}}{2}$ \\
$E_{\gamma}$ &=& $1.884 m =\frac{m_{2}+m_{3}}{2}$ \\
$E_{\delta}$ &=& $2.175 m =\frac{m_{3}+m_{5}}{2}$ \\
$E_{\varepsilon}$ &=& $2.971 m =\frac{m_{2}+m_{3}+m_{5}}{2}$ \\
\end{tabular}
\end{center}
where the mass parameter $m$ is defined by $m_{1}^{2}=\left(3-\sqrt{3}\right)m^{2}$. Levels $\beta$ and $\gamma$
are degenerate, due to the equality $m_{3}=m_{5}$; they correspond respectively to the states
$\beta_{1},\beta_{2}$ and $\gamma_{1},\gamma_{2}$, with
\begin{center}
\begin{tabular}{cclc|cclc|}
$K_{b}^{\beta_{1}}(\theta)$ &=& $S_{b3}(\theta)K_{b}^{0}(\theta)$ && $K_{b}^{\gamma_{1}}(\theta)$ &=& $S_{b2}(\theta)S_{b5}(\theta)K_{b}^{0}(\theta)$ \\
$K_{b}^{\beta_{2}}(\theta)$ &=& $S_{b5}(\theta)K_{b}^{0}(\theta)$ && $K_{b}^{\gamma_{2}}(\theta)$ &=& $S_{b2}(\theta)S_{b3}(\theta)K_{b}^{0}(\theta)$ \\
\end{tabular}
\end{center}

We list the \lq\lq fusing angles\rq\rq \, following the conventions of \cite{fring2}:

\begin{center}
\begin{tabular}{|c|c|c|c|c|c|c|c|c|}\hline
 $ a\setminus\mu$  &  $ 0$  &  $ \alpha$ & $ \beta_{1}$ & $ \beta_{2}$ &  $ \gamma_{1}$ & $\gamma_{2}$ & $\delta$ &$\varepsilon$  \\ \hline
 1 & $1^{\beta_{1}}$ & $1^{\gamma_{2}}5^{\beta_{1}}$& $3^{\gamma_{1}}$& $1^{\delta}$& $1^{\varepsilon}5_{3}^{\delta}$& & &  \\ \hline
 2 & $4^{\alpha}$ & $2^{\delta}6^{\alpha}$ & $4_{3}^{\gamma_{2}}6^{\beta_{1}}$& $4_{3}^{\gamma_{1}}6^{\beta_{2}}$& $6_{3}^{\gamma_{1}}$& $6_{3}^{\gamma_{2}}$& $4_{5}^{\varepsilon}6_{3}^{\delta}$& $6_{5}^{\varepsilon}$\\ \hline
 3 & $2^{\gamma_{1}}4^{\beta_{2}}$ & $4_{3}^{\gamma_{1}}$ & $2_{3}^{\varepsilon}4_{3}^{\delta}$ & $6_{3}^{\beta_{1}}$ & $6_{5}^{\gamma_{2}}$ & $4_{5}^{\varepsilon}$ & & \\ \hline
 4 & $1^{\varepsilon}3_{3}^{\delta}5^{\alpha}$& $3_{5}^{\varepsilon}$ & $5_{5}^{\gamma_{2}}$ & $5_{5}^{\gamma_{1}}$& & & $5_{9}^{\varepsilon}$& \\ \hline
 5 & $2^{\gamma_{2}}4^{\beta_{1}}$ & $4_{3}^{\gamma_{2}}$ & $6_{3}^{\beta_{2}}$ & $2_{3}^{\varepsilon}4_{3}^{\delta}$   & $4_{5}^{\varepsilon}$ & $6_{5}^{\gamma_{1}}$& & \\ \hline
 6 & $1^{\beta_{2}}$ & $1^{\gamma_{1}}5^{\beta_{2}}$& $1^{\delta}$& $3^{\gamma_{2}}$& & $1^{\varepsilon}5_{3}^{\delta}$& &   \\ \hline
\end{tabular}
\end{center}

\vspace{0.5cm}

Each entry in the table indicates a fusing angle as a multiple of $\frac{i\pi}{12}$; the left column refers to the
particle type which scatters off the boundary in the state indicated in the first row. The superscript refers to
the state the boundary is changing into, and the subscript refers to the order of the pole, if multiple. As in
all the other cases we will examine, the residues have always the same sign as $B$ varies in $[0,2]$, they vanish
at the extremes of the interval and sometimes also in $B=1$.

\vspace{0.5cm}

The second solution gives the same number of boundary bound states, with energies
\begin{center}
\begin{tabular}{cclc|}
$E_{\alpha}$ &=& $0.563 m =\frac{m_{1}}{2}$ \\
$E_{\beta}$ &=& $1.126 m =\frac{m_{1}+m_{6}}{2}$ \\
$E_{\gamma}$ &=& $1.538 m =\frac{m_{4}}{2}$ \\
$E_{\delta}$ &=& $2.101 m =\frac{m_{1}+m_{4}}{2}$ \\
$E_{\varepsilon}$ &=& $2.664 m =\frac{m_{1}+m_{4}+m_{6}}{2}$ \\
\end{tabular}
\end{center}
Levels $\alpha$ and $\delta$ are degenerate, with
\begin{center}
\begin{tabular}{cclc|cclc|}
$K_{b}^{\alpha_{1}}(\theta)$ &=& $S_{b6}(\theta)K_{b}^{0}(\theta)$ && $K_{b}^{\delta_{1}}(\theta)$ &=& $S_{b4}(\theta)S_{b1}(\theta)K_{b}^{0}(\theta)$ \\
$K_{b}^{\alpha_{2}}(\theta)$ &=& $S_{b1}(\theta)K_{b}^{0}(\theta)$ && $K_{b}^{\delta_{2}}(\theta)$ &=& $S_{b4}(\theta)S_{b6}(\theta)K_{b}^{0}(\theta)$ \\
\end{tabular}
\end{center}

The \lq\lq fusing angles\rq\rq \, are:

\begin{center}
\begin{tabular}{|c|c|c|c|c|c|c|c|c|}\hline
 $ a\setminus\mu$  &  $ 0$  &  $ \alpha_{1}$ & $ \alpha_{2}$ & $ \beta$ &  $ \gamma$ & $\delta_{1}$ & $\delta_{2}$ &$\varepsilon$  \\ \hline
 1 & $4^{\alpha_{1}}$ & $6^{\alpha_{2}}$& $2^{\gamma}4^{\beta}$& $2^{\delta_{2}}$& $4_{3}^{\delta_{2}}$& $4_{3}^{\varepsilon}$& $6_{3}^{\delta_{1}}$&  \\ \hline
 2 & $1^{\gamma}3^{\beta}6^{0}$ & $1^{\delta_{2}}6^{\alpha_{1}}$ & $1^{\delta_{1}}6^{\alpha_{2}}$& $1^{\varepsilon}5_{3}^{\gamma}6^{\beta}$& $3_{3}^{\varepsilon}6^{\gamma}$& $6^{\delta_{1}}$& $6^{\delta_{2}}$& $6^{\varepsilon}$\\ \hline
 3 & $1^{\delta_{1}}5^{\alpha_{2}}$ & $1^{\varepsilon}5_{3}^{\beta}$ & $3_{3}^{\delta_{2}}$ & & $5_{5}^{\delta_{1}}$ & &$5_{7}^{\varepsilon}$ & \\ \hline
 4 & $2_{3}^{\varepsilon}4_{3}^{\gamma}6^{0}$& $4_{5}^{\delta_{2}}6_{3}^{\alpha_{1}}$ & $4_{5}^{\delta_{1}}6_{3}^{\alpha_{2}}$ & $4_{7}^{\varepsilon}6_{5}^{\beta}$& $6_{7}^{\gamma}$& $6_{9}^{\delta_{1}}$& $6_{9}^{\delta_{2}}$& $6_{11}^{\varepsilon}$\\ \hline
 5 & $1^{\delta_{2}}5^{\alpha_{1}}$ & $3_{3}^{\delta_{1}}$ & $1^{\varepsilon}5_{3}^{\beta}$ & & $5_{5}^{\delta_{2}}$ & $5_{7}^{\varepsilon}$ & & \\ \hline
 6 & $4^{\alpha_{2}}$ & $2^{\gamma}4^{\beta}$& $6^{\alpha_{1}}$& $2^{\delta_{1}}$& $4_{3}^{\delta_{1}}$& $6_{3}^{\delta_{2}}$&$4_{3}^{\varepsilon}$ &  \\ \hline
\end{tabular}
\end{center}

\vspace{0.5cm}

The two solutions have different behaviours with respect to the $Z_{3}$ symmetry of the extended Dynkin diagram.
In fact, the boundary bound states obtained from the \lq\lq minimal\rq\rq \, solution also enjoy this symmetry,
while the ones obtained from the second solution don't. This is an indication that the \lq\lq minimal\rq\rq \,
solution should correspond to the free boundary condition or to another boundary condition which preserves the
$Z_{3}$ symmetry, while in the second case an operator which breaks this symmetry lives on the boundary.

\vspace{0.5cm}

\subsection{$E_{7}$-Affine Toda Field Theory}

The bulk S-matrix has been found in \cite{E7}. Ordering the particles with increasing mass and omitting the
$\theta$-dependence, the \lq\lq minimal\rq\rq \, set of reflection amplitudes is:

\begin{center}
\begin{tabular}{cclc|}
$K_{1}$&=& ${\cal K}_{1}{\cal K}_{9}{\cal K}_{17}$  \\
$K_{2}$&=& ${\cal K}_{1}{\cal K}_{7}{\cal K}_{11} {\cal K}_{53} $\\
$K_{3}$&=& ${\cal K}_{1}{\cal K}_{5}{\cal K}_{7}{\cal K}_{9}{\cal K}_{47}{\cal K}_{13} {\cal K}_{17}$ \\
$K_{4}$&=& ${\cal K}_{1}{\cal K}_{3}{\cal K}_{7}{\cal K}_{9}{\cal K}_{45}{\cal K}_{11} {\cal K}_{15}{\cal K}_{53}$ \\
$K_{5}$&=& ${\cal K}_{1}{\cal K}_{3}{\cal K}_{5}{\cal K}_{7}{\cal K}_{43}{\cal K}_{9}^{2}{\cal K}_{11}{\cal K}_{47} {\cal K}_{13}{\cal K}_{51}{\cal K}_{17}$ \\
$K_{6}$&=& ${\cal K}_{1}{\cal K}_{3}{\cal K}_{5}^{2}{\cal K}_{7}{\cal K}_{43}{\cal K}_{9}^{2}{\cal K}_{45}{\cal K}_{11}{\cal K}_{47} {\cal K}_{13}^{2}{\cal K}_{51}{\cal K}_{17}$ \\
$K_{7}$&=& ${\cal K}_{1}{\cal K}_{3}^{2}{\cal K}_{5}^{2}{\cal K}_{41}{\cal K}_{7}^{3}{\cal K}_{43}{\cal
K}_{9}^{2}{\cal K}_{45}^{2}{\cal K}_{11}^{3}{\cal K}_{47} {\cal K}_{13}{\cal K}_{49}^{2}{\cal K}_{15}^{2}{\cal
K}_{53}$ \\
\end{tabular}
\end{center}

Fixing $E_{0}=0$ and defining $M$ the mass of the lightest particle, the corresponding energy levels are:

\begin{center}
\begin{tabular}{cclc|cclc|}
$E_{\alpha}$ &=& $1.266 M =\frac{m_{5}}{2}$ && $E_{\varepsilon}$ &=&2.706 $M =\frac{m_{1}+m_{3}+m_{5}}{2}$ \\
$E_{\beta}$ &=& $1.440 M =\frac{m_{1}+m_{3}}{2}$ && $E_{\sigma}$ &=& $3.206 M =\frac{m_{1}+m_{5}+m_{6}}{2}$ \\
$E_{\gamma}$ &=& $1.940 M =\frac{m_{1}+m_{6}}{2}$ && $E_{\tau}$ &=& $3.645 M =\frac{m_{3}+m_{5}+m_{6}}{2}$ \\
$E_{\delta}$ &=& $2.380 M =\frac{m_{3}+m_{6}}{2}$ && \\
\end{tabular}
\end{center}

\vspace{0.5cm}

We list now the \lq\lq fusing angles\rq\rq \, as multiples of $\frac{i\pi}{18}$:

\begin{center}
\begin{tabular}{|c|c|c|c|c|c|c|c|c|}\hline
 $ a\setminus\mu$  &  $ 0^{+}$  &  $ \alpha^{+}$ &  $ \beta^{-}$ &  $ \gamma^{+}$& $\delta^{+}$ & $\varepsilon^{-}$ & $\sigma^{+}$ & $\tau^{+}$\\ \hline
 $1^{-}$ &   &  $8^{\beta}$ &  $2^{\delta}6^{\gamma}$ &  $4^{\varepsilon}$ &  & $2^{\tau}6_{3}^{\sigma}$ &  & \\ \hline
  $2^{+}$ & $1^{\alpha}$  & $3^{\delta}9^{\alpha}$ & $1^{\varepsilon}9^{\beta}$ & $1^{\sigma}7_{3}^{\delta}9^{\gamma}$ & $1^{\tau}5_{3}^{\sigma}9_{3}^{\delta}$ & $9_{3}^{\varepsilon}$ & $7_{5}^{\tau}9_{3}^{\sigma}$ & $9_{5}^{\tau}$\\ \hline
  $3^{-}$ & $4^{\beta}$   & $4_{3}^{\varepsilon}$ & $2^{\sigma}6_{3}^{\delta}$ & & $8_{5}^{\varepsilon}$ & $6_{5}^{\tau}$ &  & \\ \hline
  $4^{+}$ & $1^{\gamma}5^{\alpha}$ &  $1^{\sigma}7_{3}^{\gamma}9^{\alpha}$ & $5_{3}^{\varepsilon}9_{3}^{\beta}$ & $3_{3}^{\tau}5_{3}^{\sigma}9_{5}^{\gamma}$ & $5_{5}^{\tau}9^{\delta}$ & $9_{5}^{\varepsilon}$ & $9_{7}^{\sigma}$ & $9_{7}^{\tau}$\\ \hline
  $5^{+}$ & $2^{\delta}4^{\gamma}6^{\alpha}$ &  $2_{3}^{\tau}4_{3}^{\sigma}$ & $6_{5}^{\varepsilon}$ & $6_{7}^{\sigma}8_{5}^{\delta}$ & $6_{7}^{\tau}$ &  & $8_{9}^{\tau}$ & \\ \hline
  $6^{-}$ & $2^{\varepsilon}6^{\beta}$    & $6_{5}^{\varepsilon}$ & $4_{5}^{\tau}8_{5}^{\gamma}$ &  & & $8_{9}^{\sigma}$ &  & \\ \hline
  $7^{+}$ & $1^{\tau}3_{3}^{\sigma}5_{3}^{\delta}7^{\alpha}$  & $5_{7}^{\tau}9_{5}^{\alpha}$ & $7_{7}^{\varepsilon}9_{5}^{\beta}$ & $7_{9}^{\sigma}9_{7}^{\gamma}$ & $7_{11}^{\tau}9_{9}^{\delta}$ & $9_{11}^{\varepsilon}$ & $9_{13}^{\sigma}$ & $9_{15}^{\tau}$\\ \hline

  \end{tabular}
\end{center}

\vspace{0.5cm}

The signs refer to the $Z_{2}$ symmetry of the extended Dynkin diagram, choosing the convention in which the
boundary ground state is even. All the poles in the reflection amplitudes are consistent with the change of parity
induced in the boundary by the particles which create the bound states, so that this solution corresponds to a
boundary condition which preserves parity.

\vspace {0.5cm}

Starting from the second solution, we obtain the same number of boundary bound states, and energy levels:

\begin{center}
\begin{tabular}{cclc|cclc|}
$E_{\alpha}$ &=& $0.643 M =\frac{m_{2}}{2}$ && $E_{\varepsilon}$ &=& $2.494 M =\frac{m_{2}+m_{7}}{2}$ \\
$E_{\beta}$ &=& $0.985 M =\frac{m_{4}}{2}$ && $E_{\sigma}$ &=& $2.836 M =\frac{m_{4}+m_{7}}{2}$ \\
$E_{\gamma}$ &=& $1.627 M =\frac{m_{2}+m_{4}}{2}$ && $E_{\tau}$ &=& $3.478 M =\frac{m_{2}+m_{4}+m_{7}}{2}$ \\
$E_{\delta}$ &=& $1.851 M =\frac{m_{7}}{2}$ && \\
\end{tabular}
\end{center}

\vspace{0.5cm}

The \lq\lq fusing angles\rq\rq \, are:

\begin{center}
\begin{tabular}{|c|c|c|c|c|c|c|c|c|}\hline
 $ a\setminus\mu$  &  $ 0$  &  $ \alpha$ &  $ \beta$ &  $ \gamma$& $\delta$ & $\varepsilon$ & $\sigma$ & $\tau$ \\ \hline
 $1^{-}$ & $1^{\beta}5^{\alpha}9^{0}$  &  $1^{\gamma}7^{\beta}9^{\alpha}$ &  $3^{\delta}5^{\gamma}9_{3}^{\beta}$ &  $3^{\varepsilon}9_{3}^{\gamma}$ & $1^{\sigma}5_{3}^{\varepsilon}9_{3}^{\delta}$ & $1^{\tau}7_{3}^{\sigma}9_{3}^{\varepsilon}$ & $5_{3}^{\tau}9_{5}^{\sigma}$ & $9_{5}^{\tau}$\\ \hline
 $2^{+}$ & $4^{\beta}6^{\alpha}9^{0}$  & $2^{\delta}4^{\gamma}9^{\alpha}$ & $6_{3}^{\gamma}9^{\beta}$ & $2^{\sigma}8_{3}^{\delta}9^{\gamma}$ &  $4_{3}^{\sigma}6_{3}^{\varepsilon}9^{\delta}$ & $4_{3}^{\tau}9^{\varepsilon}$ & $6_{5}^{\tau}9^{\sigma}$ & $9^{\tau}$\\ \hline
 $3^{-}$ & $1^{\delta}3^{\gamma}7^{\alpha}9^{0}$   & $1^{\varepsilon}5^{\delta}9_{3}^{\alpha}$ & $1^{\sigma}7_{3}^{\gamma}9_{3}^{\beta}$ & $1^{\tau}5_{5}^{\sigma}9_{5}^{\gamma}$& $3_{3}^{\tau}7_{5}^{\varepsilon}9_{5}^{\delta}$ & $9_{7}^{\varepsilon}$ & $7_{7}^{\tau}9_{7}^{\sigma}$ & $9_{9}^{\tau}$\\ \hline
 $4^{+}$ & $2^{\delta}6^{\beta}9^{0}$ &  $2^{\varepsilon}6_{3}^{\gamma}8_{3}^{\beta}9^{\alpha}$ & $2_{3}^{\sigma}4_{3}^{\varepsilon}9^{\beta}$ & $2_{3}^{\tau}9^{\gamma}$ & $6_{5}^{\sigma}9^{\delta}$ & $6_{7}^{\tau}8_{7}^{\sigma}9^{\varepsilon}$ & $9^{\sigma}$ & $9^{\tau}$\\ \hline
 $5^{+}$ & $1^{\varepsilon}5_{3}^{\gamma}9^{0}$ &  $3_{3}^{\sigma}9_{3}^{\alpha}$ & $1^{\tau}7_{5}^{\delta}9_{3}^{\beta}$ & $7_{7}^{\varepsilon}9_{5}^{\gamma}$ & $9_{7}^{\delta}$ & $9_{9}^{\varepsilon}$ & $9_{9}^{\sigma}$ & $9_{11}^{\tau}$\\ \hline
 $6^{-}$ & $1^{\sigma}3_{3}^{\varepsilon}5_{3}^{\delta}7_{3}^{\beta}9^{0}$ & $1^{\tau}5_{5}^{\varepsilon}7_{5}^{\gamma}9_{3}^{\alpha}$ & $3_{5}^{\tau}5_{5}^{\sigma}9_{5}^{\beta}$ & $5_{7}^{\tau}9_{7}^{\gamma}$ & $7_{9}^{\sigma}9_{7}^{\delta}$ & $7_{11}^{\tau}9_{9}^{\varepsilon}$ & $9_{11}^{\sigma}$ & $9_{13}^{\tau}$\\ \hline
 $7^{+}$ & $2_{3}^{\tau}4_{5}^{\sigma}6_{5}^{\delta}8_{3}^{\alpha}9^{0}$  & $4_{7}^{\tau}6_{7}^{\varepsilon}9^{\alpha}$ & $6_{9}^{\sigma}8_{7}^{\gamma}9^{\beta}$ & $6_{11}^{\tau}9^{\gamma}$ & $8_{11}^{\varepsilon}9^{\delta}$ & $9^{\varepsilon}$ & $8_{15}^{\tau}9^{\sigma}$ & $9^{\tau}$\\ \hline

  \end{tabular}
\end{center}

\vspace{0.5cm}

In this case the boundary bound states don't have a definite parity, and this again signals the presence on the
boundary of an operator which breaks this symmetry.

\vspace{0.5cm}

\subsection{$E_{8}$-Affine Toda Field Theory}

The bulk S-matrix has been found in \cite{E8}. Ordering the particles with increasing mass and omitting the
$\theta$-dependence, the \lq\lq minimal\rq\rq \, set of reflection amplitudes is
\begin{center}
\begin{tabular}{l}
$K_{1}={\cal K}_{1}{\cal K}_{11}{\cal K}_{19}{\cal K}_{89}$  \\
$K_{2}={\cal K}_{1}{\cal K}_{7}{\cal K}_{11} {\cal K}_{13}{\cal K}_{77}{\cal K}_{19}{\cal K}_{23}{\cal K}_{89} $\\
$K_{3}={\cal K}_{1}{\cal K}_{3}{\cal K}_{9}{\cal K}_{11}{\cal K}_{71}{\cal K}_{13} {\cal K}_{17}{\cal K}_{19}
                     {\cal K}_{79}{\cal K}_{21}{\cal K}_{87}{\cal K}_{29}$ \\
$K_{4}={\cal K}_{1}{\cal K}_{5}{\cal K}_{7}{\cal K}_{9}{\cal K}_{11}{\cal K}_{71}{\cal K}_{13} {\cal K}_{15}
                     {\cal K}_{75}{\cal K}_{17}{\cal K}_{19}{\cal K}_{79}{\cal K}_{21}{\cal K}_{83}{\cal K}_{25}{\cal K}_{29}$ \\
$K_{5}={\cal K}_{1}{\cal K}_{3}{\cal K}_{5}{\cal K}_{7}{\cal K}_{9}{\cal K}_{69}{\cal K}_{11}^{2}{\cal K}_{71}
                     {\cal K}_{13}{\cal K}_{73}{\cal K}_{15}^{2}{\cal K}_{17}{\cal K}_{77}{\cal K}_{19}^{2}{\cal K}_{79}
                     {\cal K}_{21}{\cal K}_{81}{\cal K}_{23}{\cal K}_{85}{\cal K}_{27}{\cal K}_{89} $ \\
$K_{6}={\cal K}_{1}{\cal K}_{3}{\cal K}_{5}{\cal K}_{7}{\cal K}_{67}{\cal K}_{9}^{2}{\cal K}_{11}^{2}
                     {\cal K}_{71} {\cal K}_{13}^{2}{\cal K}_{73}{\cal K}_{15}{\cal K}_{75}{\cal K}_{17}^{2}{\cal K}_{77}
                     {\cal K}_{19}{\cal K}_{79}^{2}{\cal K}_{21}^{2}{\cal K}_{23}{\cal K}_{83}{\cal K}_{25}{\cal K}_{87}{\cal K}_{29}$ \\
$K_{7}={\cal K}_{1}{\cal K}_{3}{\cal K}_{5}^{2}{\cal K}_{7}^{2}{\cal K}_{67}{\cal K}_{9}^{2}{\cal K}_{69}
                     {\cal K}_{11}^{2}
                     {\cal K}_{71}^{2} {\cal K}_{13}^{3}{\cal K}_{73}{\cal K}_{15}^{2}{\cal K}_{75}^{2}{\cal K}_{17}^{3}
                     {\cal K}_{77}
                     {\cal K}_{19}^{2}{\cal K}_{79}^{2}{\cal K}_{21}^{2}{\cal K}_{81}{\cal K}_{23}{\cal K}_{83}^{2}
                     {\cal K}_{25}^{2}{\cal K}_{87}
                     {\cal K}_{29}$ \\
$K_{8}={\cal K}_{1}{\cal K}_{3}^{2}{\cal K}_{5}^{2}{\cal K}_{65}{\cal K}_{7}^{3}{\cal K}_{67}{\cal
K}_{9}^{3}{\cal K}_{69}^{2}{\cal K}_{11}^{4}
                     {\cal K}_{71}^{2} {\cal K}_{13}^{3}{\cal K}_{73}^{3}{\cal K}_{15}^{4}{\cal K}_{75}^{2}{\cal K}_{17}^{3}
                     {\cal K}_{77}^{3}
                     {\cal K}_{19}^{4}{\cal K}_{79}^{2}{\cal K}_{21}^{2}{\cal K}_{81}^{3}{\cal K}_{23}^{3}{\cal K}_{83}
                     {\cal K}_{25}{\cal K}_{85}^{2}{\cal K}_{27}^{2}{\cal K}_{89}$ \\

\end{tabular}
\end{center}

\vspace{0.5cm}

Fixing $E_{0}=0$ and defining $M$ the mass of the lightest particle, the corresponding energy levels are:

\begin{center}
\begin{tabular}{cclc|cclc}
$E_{\alpha}$ &=& $0.9945 M =\frac{m_{3}}{2}$ && $E_{\nu}$ &=& $3.1480 M =\frac{m_{4}+m_{7}}{2}$ \\
$E_{\beta}$ &=& $1.2024 M =\frac{m_{4}}{2}$ && $E_{\rho}$ &=& $3.5547 M =\frac{m_{6}+m_{7}}{2}$ \\
$E_{\gamma}$ &=& $1.6092 M =\frac{m_{6}}{2}$ && $E_{\sigma}$ &=& $3.8061 M =\frac{m_{3}+m_{4}+m_{6}}{2}$ \\
$E_{\delta}$ &=& $1.9456 M =\frac{m_{7}}{2}$ && $E_{\tau}$ &=& $4.1425 M =\frac{m_{3}+m_{4}+m_{7}}{2}$ \\
$E_{\varepsilon}$ &=& $2.1969 M =\frac{m_{3}+m_{4}}{2}$ && $E_{\psi}$ &=& $4.5493 M =\frac{m_{3}+m_{6}+m_{7}}{2}$ \\
$E_{\kappa}$ &=& $2.6037 M =\frac{m_{3}+m_{6}}{2}$ && $E_{\omega}$ &=& $4.7572 M =\frac{m_{4}+m_{6}+m_{7}}{2}$ \\
$E_{\lambda}$ &=& $2.8116 M =\frac{m_{4}+m_{6}}{2}$ && $E_{\phi}$ &=& $5.7517 M =\frac{m_{3}+m_{4}+m_{6}+m_{7}}{2}$ \\
$E_{\mu}$ &=& $2.9401 M =\frac{m_{3}+m_{7}}{2}$ && \\
\end{tabular}
\end{center}

\vspace{0.5cm}

We list now the \lq\lq fusing angles\rq\rq \, as multiples of $\frac{i\pi}{30}$:

\begin{center}

\begin{tabular}{|c|c|c|c|c|c|}\hline
 $ a\setminus\mu$  &  $ 0$  &  $ \alpha$ &  $ \beta$ &  $ \gamma$ & $\delta$  \\ \hline
1 & $1^{\alpha}$ & $3^{\delta}13^{\beta}$ & $1^{\varepsilon}7^{\delta}11^{\gamma}15^{\beta}$&
$1^{\kappa}9^{\varepsilon}15^{\gamma}$& $1^{\mu}5^{\lambda}15^{\delta}$ \\ \hline

2 & $1^{\gamma}7^{\beta}$& $1^{\kappa}7_{3}^{\varepsilon}9^{\delta}15^{\alpha}$ &
$1^{\lambda}5^{\kappa}15^{\beta}$ & $3^{\nu}7_{3}^{\lambda}13_{3}^{\delta}15^{\gamma}$ &
$1^{\rho}7_{3}^{\nu}11_{3}^{\kappa}15_{3}^{\delta}$   \\ \hline

3 & $2^{\delta}6^{\gamma}10^{\alpha}$ & $2_{3}^{\mu}4^{\lambda}6^{\kappa}12_{3}^{\gamma}14^{\beta}$ &
$2^{\nu}6_{3}^{\lambda}10_{3}^{\varepsilon}$ & $2^{\rho}8_{3}^{\mu}10_{3}^{\kappa}$ & $6_{3}^{\rho}10_{5}^{\mu}$   \\
\hline

4 & $4^{\varepsilon}6^{\delta}8^{\gamma}10^{\beta}$ & $6_{3}^{\mu}8_{3}^{\kappa}10_{3}^{\varepsilon}$ &
$2^{\rho}6_{3}^{\nu}8_{3}^{\lambda}12_{3}^{\delta}$ & $4_{3}^{\sigma}6_{3}^{\rho}10_{5}^{\lambda}$&
$4_{3}^{\tau}8_{5}^{\rho}10_{5}^{\nu}14_{5}^{\varepsilon}$   \\
\hline

5 & $1^{\mu}3^{\lambda}7^{\varepsilon}11^{\beta}$&
$3_{3}^{\sigma}5_{3}^{\rho}11_{5}^{\varepsilon}13_{3}^{\gamma}15^{\alpha}$ & $1^{\tau}9_{5}^{\mu}15_{3}^{\beta}$ &
$1^{\psi}7_{5}^{\sigma}11_{5}^{\lambda}15_{5}^{\gamma}$ & $3_{3}^{\omega}7_{5}^{\tau}11_{7}^{\nu}15_{5}^{\delta}$    \\
\hline

6 & $2^{\nu}4^{\mu}6_{3}^{\kappa}10^{\gamma}12^{\alpha}$ & $2^{\tau}8_{5}^{\nu}10_{3}^{\kappa}$ &
$4_{3}^{\tau}6_{5}^{\sigma}10_{5}^{\lambda}12_{5}^{\varepsilon}$ &
$2_{3}^{\omega}4_{3}^{\psi}12_{7}^{\kappa}14_{5}^{\delta}$ & $6_{7}^{\psi}10_{7}^{\rho}12_{7}^{\mu}$
\\ \hline

7 & $2^{\sigma}4_{3}^{\rho}6_{3}^{\nu}8_{3}^{\kappa}10_{3}^{\delta}12^{\beta}$&
$4_{5}^{\psi}6_{5}^{\tau}10_{7}^{\mu}12_{5}^{\varepsilon}$ &
$4_{5}^{\omega}8_{7}^{\sigma}10_{7}^{\nu}14_{5}^{\gamma}$ &
$6_{7}^{\omega}10_{9}^{\rho}12_{7}^{\lambda}$ & $2_{3}^{\phi}8_{9}^{\psi}12_{9}^{\nu}$   \\
\hline

8 & $1^{\omega}3_{3}^{\psi}5_{5}^{\tau}7_{5}^{\rho}9_{5}^{\lambda}11_{3}^{\delta}13^{\alpha}$ &
$1^{\phi}7_{9}^{\psi}9_{9}^{\sigma}11_{7}^{\mu}15_{5}^{\alpha}$ &
$3^{\phi}7_{9}^{\omega}11_{9}^{\nu}13_{7}^{\varepsilon}15_{5}^{\beta}$ &
$5_{9}^{\phi}11_{11}^{\rho}13_{9}^{\kappa}15_{7}^{\gamma}$ & $9_{13}^{\omega}13_{11}^{\mu}15_{9}^{\delta}$   \\
\hline
  \end{tabular}

\begin{tabular}{|c|c|c|c|c|c|}\hline
 $ a\setminus\mu$    &$\varepsilon$  & $\kappa$ & $\lambda$ & $\mu$ & $\nu$\\ \hline
1 & $3^{\nu}7^{\mu}11_{3}^{\kappa}15^{\varepsilon}$ & $3^{\rho}13_{3}^{\lambda}15^{\kappa}$& $1^{\sigma}7_{3}^{\rho}15_{3}^{\lambda}$ & $5^{\sigma}13_{3}^{\nu}15^{\mu}$& $1^{\tau}11_{3}^{\rho}15_{3}^{\nu}$\\
\hline

2 &$1^{\sigma}9_{3}^{\nu}15_{3}^{\varepsilon}$& $3^{\tau}7_{5}^{\sigma}9_{3}^{\rho}15_{3}^{\kappa}$ & $13_{5}^{\nu}15_{3}^{\lambda}$ &$1^{\psi}7_{5}^{\tau}15_{5}^{\mu}$ & $1^{\omega}5_{3}^{\psi}11_{5}^{\sigma}15_{5}^{\nu}$ \\
\hline

3  & $2_{3}^{\tau}6_{3}^{\sigma}12_{5}^{\lambda}$ & $2_{3}^{\psi}14_{5}^{\lambda}$ & $2^{\omega}8_{5}^{\tau}10_{5}^{\sigma}$ & $4_{3}^{\omega}6_{3}^{\psi}12_{7}^{\rho}14_{5}^{\nu}$ &  $6_{5}^{\omega}10_{7}^{\tau}$ \\
\hline

4 & $2^{\psi}6_{5}^{\tau}8_{5}^{\sigma}12_{5}^{\mu}$& $6_{5}^{\psi}10_{7}^{\sigma}$& $6_{5}^{\omega}12_{7}^{\rho}$ & $8_{7}^{\psi}10_{7}^{\tau}$ &  $8_{7}^{\omega}$\\
\hline

5 & $5_{5}^{\omega}13_{7}^{\lambda}15_{5}^{\varepsilon}$& $11_{9}^{\sigma}15_{7}^{\kappa}$ & $1^{\phi}9_{9}^{\psi}15_{9}^{\lambda}$ & $3_{5}^{\phi}11_{11}^{\tau}13_{9}^{\rho}15_{7}^{\mu}$ &  $15_{9}^{\nu}$  \\
\hline

6 & $10_{7}^{\sigma}$& $2_{3}^{\phi}8_{9}^{\omega}14_{9}^{\mu}$& $4_{5}^{\phi}12_{11}^{\sigma}14_{9}^{\nu}$ &
$10_{9}^{\psi}$ & $6_{9}^{\phi}10_{11}^{\omega}12_{11}^{\tau}$
\\ \hline

7  & $4_{7}^{\phi}10_{11}^{\tau}14_{9}^{\kappa}$ & $6_{9}^{\phi}10_{13}^{\psi}12_{11}^{\sigma}$ & $10_{13}^{\omega}$ & $12_{13}^{\tau}$ & $8_{13}^{\phi}14_{13}^{\rho}$   \\
\hline

8 & $7_{13}^{\phi}11_{13}^{\tau}15_{11}^{\varepsilon}$& $11_{15}^{\psi}15_{13}^{\kappa}$ & $11_{17}^{\omega}13_{15}^{\sigma}15_{13}^{\lambda}$ & $9_{17}^{\phi}15_{15}^{\mu}$ & $13_{17}^{\tau}15_{15}^{\nu}$  \\
\hline
  \end{tabular}

\begin{tabular}{|c|c|c|c|c|c|c|}\hline
 $ a\setminus\mu$ & $\rho$ & $\sigma$ & $\tau$ & $\psi$ & $\omega$ & $\phi$\\ \hline
1 & $1^{\psi}9_{3}^{\tau}15_{3}^{\rho}$& $3^{\omega}7_{3}^{\psi}15_{3}^{\sigma}$ & $11_{5}^{\psi}15_{3}^{\tau}$ & $13_{5}^{\omega}15_{3}^{\psi}$ & $1^{\phi}15_{5}^{\omega}$ & $15_{5}^{\phi}$\\
\hline

2 & $7_{5}^{\omega}15_{5}^{\rho}$& $9_{5}^{\omega}13_{7}^{\tau}15_{5}^{\sigma}$ & $1^{\phi}15_{7}^{\tau}$ & $7_{7}^{\phi}15_{7}^{\psi}$ & $15_{7}^{\omega}$ & $15_{9}^{\phi}$\\
\hline

3 & $10_{7}^{\psi}$ & $2_{3}^{\phi}$ & $6_{5}^{\phi}12_{9}^{\omega}$ &  $14_{9}^{\omega}$ & $10_{9}^{\phi}$ &\\
\hline

4 & $4_{5}^{\phi}10_{9}^{\omega}14_{9}^{\sigma}$& $6_{7}^{\phi}12_{9}^{\psi}$& $8_{9}^{\phi}$ & $10_{9}^{\phi}$ & & \\
\hline

5 & $7_{9}^{\phi}11_{11}^{\omega}15_{11}^{\rho}$& $15_{11}^{\sigma}$& $13_{13}^{\omega}15_{11}^{\tau}$ & $11_{15}^{\phi}15_{13}^{\psi}$ & $15_{15}^{\omega}$ & $15_{17}^{\phi}$ \\
\hline

6 & $12_{13}^{\psi}$ & $14_{13}^{\tau}$ & $10_{13}^{\phi}$ && $12_{17}^{\phi}$ &
\\ \hline

7 & $12_{15}^{\omega}$& $10_{17}^{\phi}$ & $14_{17}^{\psi}$ &  $12_{19}^{\phi}$ & &   \\
\hline

8 & $13_{19}^{\psi}15_{17}^{\rho}$ & $11_{21}^{\phi}15_{19}^{\sigma}$ & $15_{21}^{\tau}$ &  $15_{23}^{\psi}$ & $13_{25}^{\phi}15_{23}^{\omega}$& $15_{29}^{\phi}$ \\
\hline
  \end{tabular}

\end{center}

\vspace {0.5cm}

Starting from the second solution, we obtain the same number of boundary bound states, and energy levels:

\begin{center}
\begin{tabular}{cclc|cclc}
$E_{\alpha}$ &=& $0.5000 M =\frac{m_{1}}{2}$ && $E_{\nu}$ &=& $2.8917 M =\frac{m_{1}+m_{8}}{2} $ \\
$E_{\beta}$ &=& $0.8090 M =\frac{m_{2}}{2}$ && $E_{\rho}$ &=& $3.2007 M =\frac{m_{2}+m_{8}}{2}$ \\
$E_{\gamma}$ &=& $1.3090 M =\frac{m_{1}+m_{2}}{2}$ && $E_{\sigma}$ &=& $3.7007 M =\frac{m_{1}+m_{2}+m_{8}}{2}$ \\
$E_{\delta}$ &=& $1.4781 M =\frac{m_{5}}{2}$ && $E_{\tau}$ &=& $3.8698 M =\frac{m_{5}+m_{8}}{2}$ \\
$E_{\varepsilon}$ &=& $1.9781 M =\frac{m_{1}+m_{5}}{2}$ && $E_{\psi}$ &=& $4.3698 M =\frac{m_{1}+m_{5}+m_{8}}{2}$ \\
$E_{\kappa}$ &=& $2.2872 M =\frac{m_{2}+m_{5}}{2}$ && $E_{\omega}$ &=& $4.6789 M =\frac{m_{2}+m_{5}+m_{8}}{2}$ \\
$E_{\lambda}$ &=& $2.3917 M =\frac{m_{8}}{2}$ && $E_{\phi}$ &=& $5.1789 M =\frac{m_{1}+m_{2}+m_{5}+m_{8}}{2}$ \\
$E_{\mu}$ &=& $2.7872 M =\frac{m_{1}+m_{2}+m_{5}}{2}$ && \\
\end{tabular}
\end{center}

\vspace{0.5cm}

The \lq\lq fusing angles\rq\rq \, are:

\begin{center}

\begin{tabular}{|c|c|c|c|c|}\hline
 $ a\setminus\mu$    &$0$  & $\alpha$ & $\beta$ & $\gamma$\\ \hline
1 & $6^{\beta}10^{\alpha}15^{0}$& $2^{\delta}6^{\gamma}12^{\beta}15^{\alpha}$ & $8^{\delta}10^{\gamma}15^{\beta}$ & $2^{\kappa}8^{\varepsilon}15^{\gamma}$\\
\hline

2 & $4^{\delta}6^{\gamma}10^{\beta}12^{\alpha}15^{0}$ & $4^{\varepsilon}10^{\gamma}15^{\alpha}$ & $2^{\lambda}4^{\kappa}12_{3}^{\gamma}15^{\beta}$ & $2^{\nu}4^{\mu}8_{3}^{\lambda}14_{3}^{\delta}15^{\gamma}$\\
\hline

3  & $1^{\varepsilon}7^{\delta}11^{\beta}15^{0}$ & $3^{\lambda}7^{\varepsilon}11_{3}^{\gamma}15^{\alpha}$ & $1^{\mu}7_{3}^{\kappa}9_{3}^{\varepsilon}15_{3}^{\beta}$& $3^{\rho}7_{3}^{\mu}15_{3}^{\gamma}$ \\
\hline

4 & $1^{\lambda}3^{\kappa}13^{\alpha}15^{0}$ & $1^{\nu}3^{\mu}7_{3}^{\kappa}11_{3}^{\delta}15_{3}^{\alpha}$ & $1^{\rho}5_{3}^{\nu}13_{3}^{\gamma}15_{3}^{\beta}$ & $1^{\sigma}11_{5}^{\kappa}15_{5}^{\gamma}$\\
\hline

5 & $2^{\nu}6_{3}^{\lambda}8_{3}^{\varepsilon}10_{3}^{\delta}15^{0}$& $4_{3}^{\rho}6_{3}^{\nu}10_{5}^{\varepsilon}14_{3}^{\beta}15^{\alpha}$ & $2^{\sigma}6_{5}^{\rho}8_{5}^{\mu}10_{5}^{\kappa}15^{\beta}$ & $6_{5}^{\sigma}10_{7}^{\mu}15^{\gamma}$\\
\hline

6  &$1^{\rho}5_{3}^{\mu}7_{3}^{\lambda}11_{3}^{\gamma}15^{0}$&
$1^{\sigma}7_{5}^{\nu}9_{5}^{\lambda}15_{3}^{\alpha}$ & $3_{3}^{\tau}7_{5}^{\rho}13_{5}^{\delta}15_{3}^{\beta}$ & $3_{3}^{\psi}7_{7}^{\sigma}9_{7}^{\rho}13_{7}^{\varepsilon}15_{5}^{\gamma}$\\
\hline

7   & $1^{\tau}3_{3}^{\sigma}7_{5}^{\nu}9_{5}^{\kappa}13_{3}^{\beta}15^{0}$& $1^{\psi}5_{5}^{\tau}9_{7}^{\mu}13_{5}^{\gamma}15_{3}^{\alpha}$ & $1^{\omega}7_{7}^{\sigma}11_{7}^{\lambda}15_{5}^{\beta}$ & $1^{\phi}5_{7}^{\omega}11_{9}^{\nu}15_{7}^{\gamma}$ \\
\hline

8  & $2_{3}^{\omega}4_{5}^{\psi}6_{7}^{\tau}8_{7}^{\rho}10_{7}^{\lambda}12_{5}^{\delta}14_{3}^{\alpha}15^{0}$ & $2_{3}^{\phi}6_{9}^{\psi}8_{9}^{\sigma}10_{9}^{\nu}12_{7}^{\varepsilon}15^{\alpha}$ & $4_{7}^{\phi}6_{9}^{\omega}10_{11}^{\rho}12_{9}^{\kappa}14_{7}^{\gamma}15^{\beta}$ & $6_{11}^{\phi}10_{13}^{\sigma}12_{11}^{\mu}15^{\gamma}$ \\
\hline
  \end{tabular}

\begin{tabular}{|c|c|c|c|c|c|}\hline
 $ a\setminus\mu$    &$\delta$  & $\varepsilon$ & $\kappa$ & $\lambda$ & $\mu$ \\ \hline
1 & $4^{\lambda}6^{\kappa}10_{3}^{\varepsilon}15^{\delta}$ & $4^{\nu}6^{\mu}12_{3}^{\kappa}15^{\varepsilon}$ & $4^{\rho}10_{3}^{\mu}14_{3}^{\lambda}15^{\kappa}$ & $6_{3}^{\rho}10_{3}^{\nu}15^{\lambda}$ & $4^{\sigma}14_{3}^{\nu}15^{\mu}$ \\
\hline

2 & $6_{3}^{\mu}10_{3}^{\kappa}12_{3}^{\varepsilon}15^{\delta}$ & $10_{3}^{\mu}15^{\varepsilon}$ & $2^{\tau}12_{5}^{\mu}15^{\kappa}$ & $4_{3}^{\tau}6_{3}^{\sigma}10_{5}^{\rho}12_{5}^{\nu}15^{\lambda}$ & $2^{\psi}8_{5}^{\tau}15^{\mu}$ \\
\hline

3 & $5_{3}^{\rho}11_{5}^{\kappa}15_{3}^{\delta}$ & $3_{3}^{\tau}5_{3}^{\sigma}11_{7}^{\mu}13_{5}^{\lambda}15_{3}^{\varepsilon}$ & $15_{5}^{\kappa}$ & $1^{\psi}7_{5}^{\tau}11_{5}^{\rho}15_{7}^{\lambda}$ & $3_{3}^{\omega}13_{7}^{\rho}15_{5}^{\mu}$ \\
\hline

4 & $1^{\tau}9_{5}^{\nu}13_{5}^{\varepsilon}15_{5}^{\delta}$ & $1^{\psi}15_{7}^{\varepsilon}$ & $1^{\omega}5_{5}^{\psi}9_{7}^{\sigma}13_{7}^{\mu}15_{7}^{\kappa}$ & $3_{3}^{\omega}13_{7}^{\nu}15_{7}^{\lambda}$ & $1^{\phi}15_{9}^{\mu}$ \\
\hline

5 & $2_{3}^{\psi}6_{5}^{\tau}12_{7}^{\lambda}15^{\delta}$ & $4_{5}^{\omega}6_{5}^{\psi}12_{9}^{\nu}14_{7}^{\kappa}15^{\varepsilon}$ & $2_{3}^{\phi}6_{7}^{\omega}12_{9}^{\rho}15^{\kappa}$ & $8_{9}^{\psi}10_{9}^{\tau}15^{\lambda}$ & $6_{7}^{\phi}12_{11}^{\sigma}15^{\mu}$ \\
\hline

6 & $1^{\omega}7_{7}^{\tau}11_{7}^{\mu}15_{7}^{\delta}$ & $1^{\phi}7_{9}^{\psi}9_{9}^{\tau}15_{9}^{\varepsilon}$
& $7_{9}^{\omega}15_{9}^{\kappa}$ & $5_{7}^{\phi}11_{11}^{\sigma}15_{9}^{\lambda}$ &
$7_{11}^{\phi}9_{11}^{\omega}15_{11}^{\mu}$
\\ \hline

7 & $3_{5}^{\phi}7_{9}^{\psi}13_{9}^{\kappa}15_{7}^{\delta}$ & $13_{11}^{\mu}15_{9}^{\varepsilon}$ & $7_{11}^{\phi}11_{13}^{\tau}15_{11}^{\kappa}$ & $9_{13}^{\omega}13_{13}^{\rho}15_{11}^{\lambda}$ & $11_{15}^{\psi}15_{13}^{\mu}$ \\
\hline

8 & $8_{13}^{\omega}10_{13}^{\tau}14_{11}^{\varepsilon}15^{\delta}$ & $8_{15}^{\phi}10_{15}^{\psi}15^{\varepsilon}$ & $10_{17}^{\omega}14_{15}^{\mu}15^{\kappa}$ & $12_{17}^{\tau}14_{15}^{\nu}15^{\lambda}$ & $10_{19}^{\phi}15^{\mu}$\\
\hline

\end{tabular}

\begin{tabular}{|c|c|c|c|c|c|c|c|}\hline
 $ a\setminus\mu$    &$\nu$  & $\rho$ & $\sigma$ & $\tau$ & $\psi$ & $\omega$ & $\phi$ \\ \hline
1 & $2^{\tau}6_{3}^{\sigma}12_{3}^{\rho}15^{\nu}$ & $8_{3}^{\tau}10_{3}^{\sigma}15^{\rho}$ & $2^{\omega}8_{3}^{\psi}15^{\sigma}$ & $6_{3}^{\omega}10_{5}^{\psi}15^{\tau}$ & $6_{3}^{\phi}12_{5}^{\omega}15^{\psi}$ & $10_{5}^{\phi}15^{\omega}$ & $15^{\phi}$\\
\hline

2 & $4_{3}^{\psi}10_{5}^{\sigma}15^{\nu}$ & $4_{3}^{\omega}12_{7}^{\sigma}15^{\rho}$ & $4_{3}^{\phi}14_{7}^{\tau}15^{\sigma}$ & $6_{5}^{\phi}10_{7}^{\omega}12_{7}^{\psi}15^{\tau}$ & $10_{7}^{\phi}15^{\psi}$ & $12_{9}^{\phi}15^{\omega}$ &$15^{\phi}$ \\
\hline

3 & $7_{5}^{\psi}11_{7}^{\sigma}15_{7}^{\nu}$ & $1^{\phi}7_{7}^{\omega}9_{7}^{\psi}15_{9}^{\rho}$ & $7_{7}^{\phi}15_{9}^{\sigma}$ & $11_{9}^{\omega}15_{9}^{\tau}$ & $11_{11}^{\phi}15_{9}^{\psi}$ & $15_{11}^{\omega}$ & $15_{11}^{\phi}$\\
\hline

4 & $3_{3}^{\phi}7_{7}^{\omega}11_{9}^{\tau}15_{9}^{\nu}$ & $13_{9}^{\sigma}15_{9}^{\rho}$ & $11_{11}^{\omega}15_{11}^{\sigma}$ & $13_{11}^{\psi}15_{11}^{\tau}$ & $15_{13}^{\psi}$ & $13_{13}^{\phi}15_{13}^{\omega}$ & $15_{15}^{\phi}$\\
\hline

5 & $10_{11}^{\psi}14_{11}^{\rho}15^{\nu}$ & $8_{11}^{\phi}10_{11}^{\omega}15^{\rho}$ & $10_{13}^{\phi}15^{\sigma}$ & $15^{\tau}$ & $14_{15}^{\omega}15^{\psi}$ &$15^{\omega}$ & $15^{\phi}$\\
\hline

6 & $15_{11}^{\nu}$ & $13_{13}^{\tau}15_{11}^{\rho}$ & $13_{15}^{\psi}15_{13}^{\sigma}$ & $11_{15}^{\phi}15_{15}^{\tau}$ & $15_{17}^{\psi}$ & $15_{17}^{\omega}$ & $15_{19}^{\phi}$ \\
\hline

7 & $9_{15}^{\phi}13_{15}^{\sigma}15_{13}^{\nu}$ & $15_{15}^{\rho}$ & $15_{17}^{\sigma}$ & $13_{19}^{\omega}15_{17}^{\tau}$ & $13_{21}^{\phi}15_{19}^{\psi}$ & $15_{21}^{\omega}$ & $15_{23}^{\phi}$ \\
\hline

8 & $12_{19}^{\psi}15^{\nu}$ & $12_{21}^{\omega}15^{\rho}$ & $12_{23}^{\phi}15^{\sigma}$ & $14_{23}^{\psi}15^{\tau}$ &$15^{\psi}$ & $14_{27}^{\phi}15^{\omega}$ & $15^{\phi}$\\
\hline

\end{tabular}

\end{center}

\newpage
\resection{Pole interpretation}

In the procedure of applying the bootstrap equation (\ref{boundst}) we have considered all the odd order poles
with positive residue. This property, however, is necessary but not sufficient for the creation of a boundary
bound state. In fact, Dorey, Tateo and Watts (\cite{tateo}) have proposed two kinds of mechanisms which can
describe some of the poles without involving new boundary bound states.

\vspace{0.5cm}

The first one is a \lq $u$-channel\rq \, mechanism, and it is invoked when an excited state reflection factor
$K_{a}^{\beta}$ has a pole at the same place $\theta_{a}=i\eta_{a\alpha}^{\beta}$ as the pole in $K_{a}^{\alpha}$
which generated $\beta$.

\vspace{5.5cm}

\begin{figure}[h]
\setlength{\unitlength}{0.0125in}
\begin{picture}(40,0)(60,470)

\put(210,470){$s$-channel}
\thicklines \put(240,490){\line(0,1){140}}

\put(240,490){\line(2,1){15}} \put(240,500){\line(2,1){15}} \put(240,520){\line(2,1){15}}
\put(240,530){\line(2,1){15}} \put(240,540){\line(2,1){15}} \put(240,550){\line(2,1){15}}
 \put(240,570){\line(2,1){15}} \put(240,580){\line(2,1){15}}
\put(240,590){\line(2,1){15}} \put(240,600){\line(2,1){15}} \put(240,620){\line(2,1){15}}

\put(240,580){\line(-1,2){25}} \put(240,540){\line(-1,-2){25}}

\put(245,510){$\alpha$}\put(245,560){$\beta$}\put(245,610){$\alpha$}

\put(205,490){$a$} \put(205,625){$a$}

\put(380,470){$u$-channel}
\thicklines \put(410,490){\line(0,1){140}}

\put(410,490){\line(2,1){15}} \put(410,500){\line(2,1){15}} \put(410,520){\line(2,1){15}}
\put(410,530){\line(2,1){15}} \put(410,540){\line(2,1){15}} \put(410,550){\line(2,1){15}}
 \put(410,570){\line(2,1){15}} \put(410,580){\line(2,1){15}}
\put(410,590){\line(2,1){15}} \put(410,600){\line(2,1){15}} \put(410,620){\line(2,1){15}}

\put(410,580){\line(-2,-5){25}} \put(410,540){\line(-2,5){25}}

\put(415,510){$\beta$}\put(415,560){$\alpha$}\put(415,610){$\beta$}

\put(375,510){$a$} \put(375,605){$a$}
\end{picture}
 \caption{First mechanism}
 \end{figure}
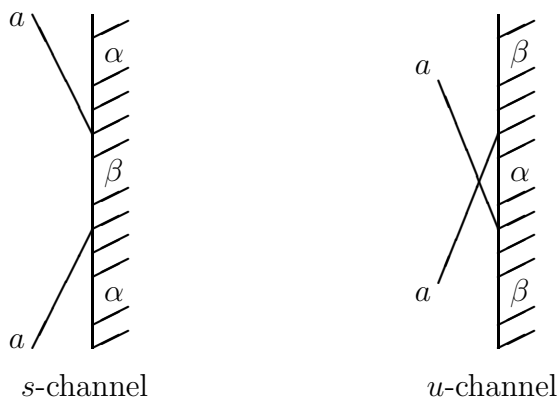

The pole in exam doesn't excite the boundary to a new bound state, but simply corresponds to going back from
$\beta$ to $\alpha$. This rule, reasonable but non properly founded, has a clear explanation in the case of a
theory with defect (\cite{defect}), where not just reflection but also transmission is allowed. In that case it
is shown with an explicit example how a pole at $\theta=i\eta$ with this property can be neglected, because,
although it has positive residue in the reflection amplitude, its residue in the transmission one is negative. At
the same time, both amplitudes have a positive residue pole at $\theta=i(\pi-\eta)$, which exactly corresponds to
going back to the original boundary state. Unfortunately, integrable defect theories seem to apply only to
quasi-free systems.

\vspace{0.5cm}

The other possibility is a boundary generalization of the Coleman-Thun mechanism: in some cases a pole can be
described by on-shell diagrams, different from the one in Figure 1, which correspond to multiple rescattering
processes. These methods have also been applied in \cite{delius} and \cite{dorey}, with the hypothesis that, if
an alternative diagram can be drawn, then the pole in exam doesn't correspond to the creation of a boundary bound
state.

In general, the order of a certain diagram is given by $P-2L$, where $P$ and $L$ are respectively the number of
propagators and loops. When dealing with a boundary, however, vertex factors can also be given by reflection
amplitudes. If these amplitudes have poles (or zeroes) at the rapidities dictated by the on-shell condition, then
their effect will be to raise (or lower) the order of the diagram.

As we have seen, the reflection amplitudes of the form (\ref{K}) never have coupling-independent zeroes in the
physical strip, hence the order of a given diagram could only be raised by their insertions. In this case, the
only two diagrams which can describe a first order pole are:

\vspace{5.5cm}

\begin{figure}[h]
\setlength{\unitlength}{0.0125in}
\begin{picture}(40,0)(60,470)

\put(190,470){Type 1}
\thicklines \put(240,490){\line(0,1){140}}

\put(240,490){\line(2,1){15}} \put(240,500){\line(2,1){15}} \put(240,510){\line(2,1){15}}
\put(240,520){\line(2,1){15}} \put(240,530){\line(2,1){15}} \put(240,540){\line(2,1){15}}
\put(240,550){\line(2,1){15}} \put(240,560){\line(2,1){15}} \put(240,570){\line(2,1){15}}
\put(240,580){\line(2,1){15}} \put(240,590){\line(2,1){15}} \put(240,600){\line(2,1){15}}
\put(240,610){\line(2,1){15}} \put(240,620){\line(2,1){15}}

\put(240,560){\line(-1,0){60}} \put(180,560){\line(-2,3){30}} \put(180,560){\line(-2,-3){30}}

\put(210,565){$b$} \put(140,515){$a$} \put(140,600){$a$}

\put(410,470){Type 2}
\thicklines \put(460,490){\line(0,1){140}}

\put(460,490){\line(2,1){15}} \put(460,500){\line(2,1){15}} \put(460,510){\line(2,1){15}}
\put(460,520){\line(2,1){15}} \put(460,530){\line(2,1){15}} \put(460,540){\line(2,1){15}}
\put(460,550){\line(2,1){15}} \put(460,560){\line(2,1){15}} \put(460,570){\line(2,1){15}}
\put(460,580){\line(2,1){15}} \put(460,590){\line(2,1){15}} \put(460,600){\line(2,1){15}}
\put(460,610){\line(2,1){15}} \put(460,620){\line(2,1){15}}

\put(460,560){\line(-2,1){60}} \put(460,560){\line(-2,-1){60}}

\put(400,530){\line(0,1){60}}

\put(400,590){\line(-2,3){30}} \put(400,530){\line(-2,-3){30}}

\put(390,560){$c$} \put(360,485){$a$} \put(360,630){$a$} \put(430,580){$b$} \put(430,535){$b$}

\end{picture}
 \caption{First order diagrams}
 \end{figure}
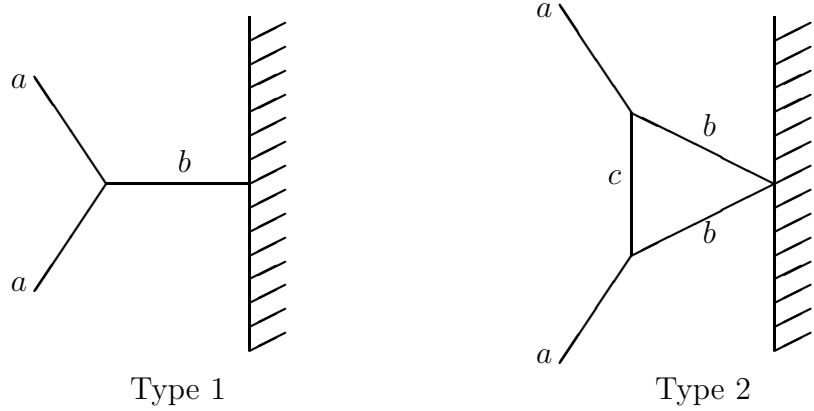

If we call $\eta_{a}$ the pole we are interested in, we can see that it is possible to draw Type 1 diagram
(already introduced in \cite{ghoszam}) if there is a particle $b$ such that $i u_{ab}^{a}=\eta_{a}+i
\frac{\pi}{2}$ and such that the corresponding reflection amplitude $K_{b}(\theta)$ has an odd order pole with
positive residue at $\theta=i\frac{\pi}{2}$. Type 2 diagram can be drawn if there are two particles $b$ and $c$
such that $u_{bc}^{a}<\frac{\pi}{2}$ and $\eta_{a}=i \left(u_{bc}^{a}+u_{ab}^{c}-\pi\right)$; the amplitude
$K_{b}(\theta)$ has to be evaluated at $\eta_{b}=i u_{bc}^{a}$.

A second order diagram, which will describe third or higher order poles of $K_{a}^{\alpha}$, is:

\vspace{5cm}

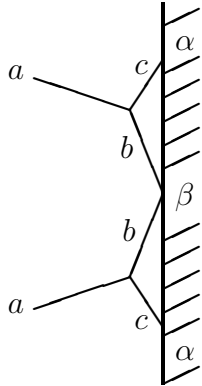
\begin{figure}[h]
\setlength{\unitlength}{0.0125in}
\begin{picture}(40,0)(60,470)
\thicklines \put(350,460){\line(0,1){160}}

\put(350,460){\line(2,1){15}} \put(350,480){\line(2,1){15}} \put(350,490){\line(2,1){15}}
\put(350,500){\line(2,1){15}} \put(350,510){\line(2,1){15}} \put(350,520){\line(2,1){15}}
\put(350,550){\line(2,1){15}} \put(350,560){\line(2,1){15}}

\put(350,570){\line(2,1){15}} \put(350,580){\line(2,1){15}} \put(350,590){\line(2,1){15}}
\put(350,610){\line(2,1){15}}

\put(350,540){\line(-2,5){14}} \put(350,540){\line(-2,-5){14}}

\put(336,575){\line(2,3){14}} \put(336,505){\line(2,-3){14}}

\put(336,575){\line(-3,1){40}} \put(336,505){\line(-3,-1){40}}

\put(355,535){$\beta$} \put(355,470){$\alpha$} \put(355,600){$\alpha$}

\put(285,588){$a$} \put(285,490){$a$} \put(332,555){$b$} \put(333,520){$b$} \put(338,590){$c$} \put(338,483){$c$}
\end{picture}
 \caption{Type 3 (second order)}
 \end{figure}

This diagram can be drawn if there are $b$ and $c$ such that $K_{c}^{\beta}$ has a pole at $\eta_{c}=i
u_{ac}^{b}-\eta_{a}$ creating the boundary bound state $\alpha$ or, if \,
$\rm{Im}\left(\eta_{c}\right)>\frac{\pi}{2}$, such that  $K_{c}^{\alpha}$ has a pole at $i\pi-\eta_{c}$ creating
$\beta$. The amplitude $K_{b}^{\beta}$ has to be evaluated at $\eta_{b}=i u_{ab}^{c}+\eta_{a}-i\pi$.

\vspace{0.5cm}

A third order diagram, with 13 propagators and 5 loops, is:

\vspace{6.5cm}

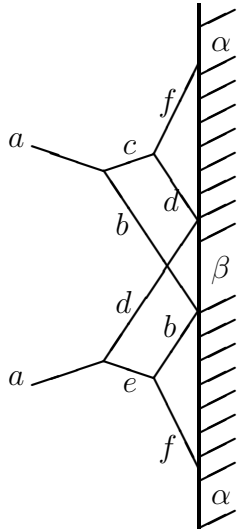
\begin{figure}[h]
\setlength{\unitlength}{0.0125in}
\begin{picture}(40,0)(60,470)
\thicklines \put(350,460){\line(0,1){220}}

\put(350,460){\line(2,1){15}} \put(350,480){\line(2,1){15}} \put(350,490){\line(2,1){15}}
\put(350,500){\line(2,1){15}} \put(350,510){\line(2,1){15}} \put(350,520){\line(2,1){15}}
\put(350,530){\line(2,1){15}} \put(350,540){\line(2,1){15}} \put(350,550){\line(2,1){15}}
\put(350,580){\line(2,1){15}}

\put(350,590){\line(2,1){15}} \put(350,600){\line(2,1){15}} \put(350,610){\line(2,1){15}}
\put(350,620){\line(2,1){15}} \put(350,630){\line(2,1){15}} \put(350,640){\line(2,1){15}}
\put(350,650){\line(2,1){15}} \put(350,670){\line(2,1){15}}
\put(355,565){$\beta$} \put(355,470){$\alpha$} \put(355,660){$\alpha$}

\put(350,550){\line(-2,3){40}} \put(350,590){\line(-2,-3){40}}

\put(310,610){\line(3,1){21}} \put(310,530){\line(3,-1){21}}

\put(331,617){\line(2,-3){19}} \put(331,617){\line(1,2){19}}

\put(331,523){\line(2,3){19}} \put(331,523){\line(1,-2){19}}

\put(310,610){\line(-3,1){30}} \put(310,530){\line(-3,-1){30}}

\put(270,620){$a$} \put(270,520){$a$}

\put(315,583){$b$} \put(318,617){$c$} \put(315,550){$d$} \put(333,635){$f$} \put(333,490){$f$}

\put(318,517){$e$}

\put(335,540){$b$} \put(335,593){$d$}
\end{picture}
 \caption{Type 4 (third order)}
 \end{figure}

This diagram can be drawn if there are all the opportune bulk fusing angles, and one has to evaluate the two
amplitudes $K_{b}^{\beta}$ and $K_{d}^{\beta}$ at the rapidities dictated by the on-shell condition.

\vspace{0.5cm}

We will now investigate if some of our poles can be described by these mechanisms. In none of the examined
reflection amplitudes \lq$u$-channel\rq \, diagrams can explain any pole. However, many generalized Coleman-Thun
diagrams can be drawn, with interesting consequences.

\vspace{0.5cm}

\subsection{Analysis of the $E_{7}$ pole structure}

\subsubsection{\lq\lq Minimal\rq\rq \, solution}

We start considering the reflection matrix in the ground state. Type 1 diagram can never be drawn (none of the
seven particles couples to the boundary at $\theta=i\frac{\pi}{2}$). If $a=1,2$, neither can type 2, lacking $b$
and $c$ such that $u_{bc}^{a}<\frac{\pi}{2}$.

However, many poles of the remaining amplitudes can be explained by this diagram; we list all the possible
corresponding choices of $b$ and $c$ in the following table:

\begin{center}
\begin{tabular}{|l|l|l|l|l|l|}\hline
\hspace{1mm} $a$  & \hspace{2mm} 3 & \hspace{2mm} 4 & \hspace{9mm} 5 & \hspace{9mm} 6 & \hspace{1.5cm} 7 \\ \hline
\hspace{1mm} $\eta_{a}$ & \hspace{2mm} $4^{\beta}$ & \hspace{2mm} $1^{\gamma}$ &\hspace{2mm} $2^{\delta}$
\hspace{6mm} $4^{\gamma}$ &\hspace{2mm} $2^{\varepsilon}$ \hspace{6mm} $6^{\beta}$ &\hspace{2mm} $1^{\tau}$
\hspace{6mm} $3_{3}^{\sigma}$ \hspace{6mm} $5_{3}^{\delta}$ \\ \hline
 $(b,c)$ & $(2,1)$& $(1,1)$ & $(1,3)$ \hspace{1mm} $(3,1)$ & $(1,5)$ \hspace{1mm} $(5,1)$ & $(1,6)$ \hspace{1mm} $(6,1)$ \hspace{1mm} $(6,3)$ \\
         &        &         &                              & $(4,1)$                      & $(2,5)$ \hspace{1mm} $(3,6)$ \\
         &        &         &                              & $(2,3)$                      &                              \\ \hline
  \end{tabular}
\end{center}

\vspace{0.5cm}

The two triple poles of $K_{7}^{0}$ are described by this diagram because, in the case $(b,c)=(6,1)$, $K_{6}^{0}$
has a double pole at $\eta_{6}=i u_{16}^{7}=i \frac{4}{18}\pi$, and in the other two cases the fusing angle
$u_{36}^{7}$ corresponds to a triple pole of the $S$-matrix.

In this way, excluding the excitation diagrams, the only boundary bound state that we can get from the ground
state is $\alpha$. If we repeat the above procedure for the $K_{b}^{\alpha}$ amplitudes, we can exclude the
creation of other boundary bound states. Here we list which poles are explained by the three kinds of diagram
seen:

\begin{center}
\begin{tabular}{|l|l|l|l|l|l|l|l|}\hline
\hspace{1mm} $a$  & \hspace{1mm} 1                 & \hspace{1mm} 2           & \hspace{2mm} 3 & \hspace{9mm}
4                                         & \hspace{9mm} 5                                           &
\hspace{2mm} 6        & \hspace{2mm} 7 \\ \hline \hspace{1mm} $\eta_{a}$ & \hspace{1mm} $8^{\beta}$ &\hspace{1mm}
$3^{\delta}$ &\hspace{2mm} $4_{3}^{\varepsilon}$ &\hspace{2mm} $1^{\sigma}$ \hspace{6mm} $7_{3}^{\gamma}$
&\hspace{2mm} $2_{3}^{\tau}$ \hspace{6mm} $4_{3}^{\sigma}$ & \hspace{2mm} $6_{5}^{\varepsilon}$ & \hspace{2mm}
$5_{7}^{\tau}$ \\ \hline Type                    & \hspace{1mm} 1            & \hspace{1mm} 1 &\hspace{2mm}
2                     & \hspace{2mm} 2           \hspace{8mm} 3                &\hspace{2mm} 2 \hspace{8mm}
2                             & \hspace{2mm} 2        & \hspace{2mm} 3 \\ \hline
 $(b,c)$ & $(4)$                       & $(2)$       & $(2,1)$                           & $(1,1)$                  \hspace{1mm} $(4,4)$          & $(1,3)$ \hspace{1mm} $(3,1)$                             & $(5,1)$               & $(7,5)$        \\ \hline

\end{tabular}
\end{center}

\vspace{0.5cm}

All the type 3 diagrams mentioned have $\alpha$ as external boundary state, and the ground state as intermediate
one. Poles of high order can be described by many different diagrams, but we have listed just one of the possible
choices. As an example, the  triple pole at $\eta_{5}=i\frac{2}{18}\pi$ admits also a description in terms of
type 3 diagram with $(b,c)=(7,2)$. The order seven for the pole at $\eta_{7}=i\frac{5}{18}\pi$ is due to the facts
that $K_{7}^{0}$ has a simple pole at $\eta_{7}=i\frac{1}{18}\pi$, and $u_{57}^{5}$ corresponds to a fifth-order
pole of the $S$-matrix.

\subsubsection{Second solution}

As we will see, in this case the above mechanisms cannot explain a number of poles sufficient to exclude the
existence of some boundary bound states.

Let's start from the ground state. Type 1 diagram cannot be drawn, because there are not the appropriate bulk
fusing angles. If $a=1,2$, we already know that neither can type 2, but for the remaining particles it explains
the following poles:

\begin{center}
\begin{tabular}{|l|l|l|l|l|l|}\hline
\hspace{1mm} $a$        & \hspace{2mm} 3            & \hspace{9mm}
5                                               & \hspace{9mm} 6                                              &
\hspace{9mm} 7 \\ \hline \hspace{1mm} $\eta_{a}$ & \hspace{2mm} $3^{\gamma}$ & \hspace{2mm} $1^{\varepsilon}$
\hspace{6mm} $5_{3}^{\gamma}$ &\hspace{2mm} $1^{\sigma}$ \hspace{6mm} $3_{3}^{\varepsilon}$ &\hspace{2mm}
$2_{3}^{\tau}$ \hspace{6mm} $4_{5}^{\sigma}$  \\ \hline
 $(b,c)$                & $(1,2)$                   & $(2,2)$ \hspace{1mm} $(4,2)$                                 & $(1,4)$ \hspace{1mm} $(3,2)$                                & $(4,4)$ \hspace{1mm} $(5,4)$  \\ \hline

\end{tabular}
\end{center}

In this way, from the ground state we get the excited stated $\alpha$, $\beta$, $\delta$. It is now easy to see
that in the corresponding amplitudes there are simple poles which cannot be explained with alternative diagrams,
and which excite the boundary to all the other bound states $\gamma$, $\varepsilon$, $\sigma$, $\tau$. These are

\begin{center}
\begin{tabular}{|c|c|c|c|c|}\hline
 $K_{a}^{\mu}$ & $K_{1}^{\alpha}$ & $K_{3}^{\alpha}$  & $K_{1}^{\delta}$ & $K_{1}^{\varepsilon}$\\ \hline
 $\eta_{a}$    & $1^{\gamma}$     & $1^{\varepsilon}$ & $1^{\sigma}$     & $1^{\tau}$\\ \hline

\end{tabular}
\end{center}

\vspace{0.5cm}

\subsection{The same analysis for $E_{6}$ and $E_{8}$}

In these cases, none of the four solutions admits a reduction of the boundary bound states number by means of
generalized Coleman-Thun diagrams.

In the $E_{6}$ case, this can be easily seen if we notice that particles $1$, $2$ and $6$, for which Type 2
diagram is not allowed, have simple poles at rapidities which forbid also Type 1 diagram, and are able to generate
the whole set of boundary bound states.

An analogous mechanism works in the $E_{8}$ case, because Type 2 diagram, when applied to light particles,
describes a narrow range of possible poles.

\vspace{1cm}

\resection{Perturbed Minimal Models}

It is now interesting to see if the results obtained for affine Toda field theories can be extended to minimal
models perturbations.

It is known that the minimal parts of the Toda $S$-matrices, defined as
\begin{equation}
S_{ab}^{min}(\theta)=\prod_{x}s_{\frac{x+1}{h}}(\theta)s_{\frac{x-1}{h}}(\theta),
\end{equation}
correspond to the scattering amplitudes of certain perturbed conformal field theories. The $E_{6}$, $E_{7}$ and
$E_{8}$ Toda theories are related respectively to the thermal perturbation of the tricritical 3-state Potts model,
the thermal perturbation of the tricritical Ising model and the magnetic perturbation of the Ising model.

In the bulk theory, \lq\lq dressing\rq\rq \, a minimal $S$-matrix with coupling constant-dependent CDD-factors
doesn't induce any change in the bound states spectrum. These factors, in fact, don't introduce new poles in the
physical strip, and don't alter the sign of the existing ones' residues.

Starting from a Toda $S$-matrix of the form (\ref{def}), where $B$ is real and varies in $[0,2]$, we can recover
its minimal part performing the so-called \lq\lq roaming limit\rq\rq, which consists in taking $B=1+iC$ and
sending the real quantity $C$ to infinity.

In general, the residue of a given pole is a real function of $C$ which preserves the same sign it had for real
$B$; its limit as $C$ tends to infinity is the value dictated by the minimal $S$-matrix.

\vspace{0.5cm}

Reflection matrices of the form (\ref{K}) are manifestly factorized in a minimal part, which satisfies equations
(\ref{unit})-(\ref{boot}) with the minimal $S$-matrix, and a set of coupling constant-dependent factors, which
admit a \lq\lq roaming limit\rq\rq \, of the same form as for the $\{x\}_{\theta}$ blocks. We already know from
(\ref{poles}) that these factors don't introduce new poles in the physical strip $0\leq
Im\theta\leq\frac{\pi}{2}$.

As we have seen in eq.(\ref{colour}), a general reflection amplitude is a product of the two kinds of blocks
$\{x\}_{\theta}$ and ${\cal K}_{y}(\theta)$. Let us assume that ${\cal K}_{y}(\theta)$ has a pole at
$\theta_{0}=i\frac{\pi}{h}\eta$; the CDD-factors of the block $\{x\}_{\theta}$, evaluated at that rapidity, give:
\begin{equation}
\frac{1}{s_{\frac{x+1-B}{h}}(\theta_{0})s_{\frac{x-1+B}{h}}(\theta_{0})}=\frac{\sin\left[\frac{\pi}{2h}(\eta-x-1+B)\right]\sin\left[\frac{\pi}{2h}(\eta-x+1-B)\right]}{\sin\left[\frac{\pi}{2h}(\eta+x+1-B)\right]\sin\left[\frac{\pi}{2h}(\eta+x-1+B)\right]}.
\end{equation}
If $\eta\neq x$, this quantity is always positive, while if $\eta=x$ we have
\begin{equation}\label{neg}
\frac{1}{s_{\frac{x+1-B}{h}}(\theta_{0})s_{\frac{x-1+B}{h}}(\theta_{0})}=-\frac{\sin^{2}\left[\frac{\pi}{2h}(1-B)\right]}{\sin\left[\frac{\pi}{2h}(2x+1-B)\right]\sin\left[\frac{\pi}{2h}(2x-1+B)\right]},
\end{equation}
which is always negative as $B$ varies in $[0,2]$, and vanishes in $B=1$. If we now parameterize $B=1+iC$, as a
function of $C$ we have
\begin{equation}\label{neg}
\frac{1}{s_{\frac{x+1-B}{h}}(\theta_{0})s_{\frac{x-1+B}{h}}(\theta_{0})}=\frac{\cos\left[\frac{\pi}{h}(\eta-x)\right]-\cosh\left(\frac{\pi}{h}C\right)}{\cos\left[\frac{\pi}{h}(\eta+x)\right]-\cosh\left(\frac{\pi}{h}C\right)},
\end{equation}
which is a positive quantity for every $C$, and if $\eta=x$ vanishes in $C=0$. This means that, in the presence of
a block $\{\eta\}$, the corresponding pole can have different signs whether we are considering the minimal or the
Toda reflection amplitude.

This phenomenon is not present in the bulk theory, because $S$-matrix poles are always located at positions
shifted by $\pm 1$ with respect to the blocks parameters $x$.

An analogous behaviour is produced by the CDD-factors of a block ${\cal K}_{y}(\theta)$, evaluated at
$\theta_{0}=i\frac{\pi}{h}\eta$ with $\eta=\frac{3h-y}{2}$. This, however, never happens for the $E_{n}$ series
elements, because all the parameters $y$ are odd, but the poles are always located at entire multiples of
$i\frac{\pi}{h}$.

\vspace{0.5cm}

From this analysis we can conclude that a reflection amplitude pole located at $\theta_{0}=i\frac{\pi}{h}\eta$
will have a residue with different sign in the minimal and in the Toda theory if this reflection amplitude has an
odd number of $\{\eta\}$ blocks.

If we think to the bulk situation this new possibility seems problematic. We have to remember, however, that the
two integrable field theories defined by a Toda Lagrangian and a minimal model perturbation correspond in the UV
limit to completely different conformal field theories. Hence, although they share the minimal part of the
$S$-matrix, they could be governed by very different integrable boundary conditions, and this might become
manifest in distinct bound states structures of the corresponding reflection amplitudes.

\vspace{0.5cm}

We will now study this phenomenon in the three examined theories.

\subsection{The Tricritical Ising Model}

Analyzing the \lq\lq minimal\rq\rq \, solution, we find many situations of the type described above, always
corresponding to poles with negative residue in the Toda theory, and positive in the minimal one. We list the
additional \lq\lq fusing angles\rq\rq \, with the usual conventions:

\begin{center}
\begin{tabular}{|c|c|c|c|c|c|c|c|c|}\hline
 $ a\setminus\mu$  &  $ 0$  &  $ \alpha$ &  $ \beta$ &  $ \gamma$& $\delta$ & $\varepsilon$ & $\sigma$ & $\tau$\\ \hline
  $1$ &   &   &   &   &  &  &  & \\ \hline
  $2$ &   &   &   &   &  &  &  & \\ \hline
  $3$ &   & $3$ &  & $3$ & $3$ & $3$ &  & \\ \hline
  $4$ &   &  $4$ & $4$  & $4$  &  &  &  & $4$ \\ \hline
  $5$ &   &  $1,\:3,\:5$  & $3$ & $3,\:5$ & $5$ & $1,\:5$ & $1$ & $1,\:3$ \\ \hline
  $6$ &   & $3$ &  & $1$ & $1$ & $3$ & $1,\:3$ & $1,\:3$\\ \hline
  $7$ &   &  &  & $4_{3}$ & $4_{3}$ &  & $4_{3}$ & $4_{3}$\\ \hline

  \end{tabular}
\end{center}

\vspace{0.5cm}

This poles create twenty new different boundary bound states, with reflection amplitudes of the form
\begin{equation}\label{noncolour}
K_{a}^{\mu}\left(\theta\right)=\left(\prod_{b}S_{ab}^{\pm
1}\left(\theta\right)\right)K_{a}^{0}\left(\theta\right),
\end{equation}
where the $b$'s can now have different colours with respect of the bicolouration of the Dynkin diagram, and the
corresponding energy levels are related by
\begin{equation}\label{level}
E_{\mu}=E_{0}+\frac{1}{2}\sum_{b}\left(\pm m_{b}\right).
\end{equation}

If we analyze the reflection amplitudes on this new boundary states, we can see that their poles generate a
cascade of other bound states, with the possibility to have
\begin{equation}\label{noncolour}
K_{a}^{\mu}\left(\theta\right)=\left(\prod_{b}S_{ab}^{\pm
n}\left(\theta\right)\right)K_{a}^{0}\left(\theta\right), \qquad E_{\mu}=E_{0}+\frac{1}{2}\sum_{b}\left(\pm
n\,m_{b}\right),
\end{equation}
with $n=1,2,...$\,.

This seems an indication that in this case the bootstrap doesn't close on a finite number of boundary states.

For the states examined we have checked that the mentioned \lq$u$-channel\rq \, mechanism cannot be applied to any
new pole. However, if we consider the generalized Coleman-Thun diagrams we can explain all the new poles
introduced on the eight Toda states by the \lq\lq roaming\rq\rq \, limit. This seems an amazing coincidence,
because we can always use Type 2 diagram, and exactly particles $1$ and $2$, for which this diagram can't be
drawn, don't produce any new pole. In this way we are also left with the two possibilities of a bootstrap closing
on two or eight boundary bound states.

\vspace{0.5cm}

Essentially the same situation arises with the second solution: the \lq\lq roaming\rq\rq \, limit introduces many
new positive residue poles in a similar way, and again Coleman-Thun diagrams can describe all of them. This time
also particles $1$ and $2$ generate new states, but we can use both Type 1 and Type 2 diagrams.

\subsection{The same analysis for $E_{6}$ and $E_{8}$}

Also in the $E_{8}$ case the \lq\lq roaming\rq\rq \, limit produces many additional positive residue poles, which
seem to indicate a non-closing bootstrap, but again we have an almost incredible coincidence between the new
poles and the ones we can explain with Coleman-Thun diagrams, so that for both solutions these mechanisms allow
the bootstrap to close on sixteen states.

\vspace{0.5cm}

In the $E_{6}$ case the situation is more delicate. As before, the \lq\lq roaming\rq\rq \, limit introduces many
new poles, and again the bootstrap presumably doesn't close. The problem is that now the Coleman-Thun diagrams
can describe almost all these poles, but in both solutions four of them remain unexplained. For the \lq\lq
minimal\rq\rq \, solution these are simple poles located at $\theta=i\frac{3}{12}\pi$ in $K_{2}^{\mu}$ with
$\mu=\beta_{1},\beta_{2},\gamma_{1},\gamma_{2}$, and they generate a cascade of states. In the second solution
case these poles, located at $\theta=i\frac{3}{12}\pi$ in $K_{4}^{\mu}$ with
$\mu=\alpha_{1},\alpha_{2},\delta_{1},\delta_{2}$, are triple, so that it is more difficult to conclude that they
don't admit alternative diagrams. However, we have tried to explain them with all kinds of diagram mentioned
(including Type 4), but we haven't succeeded.

This seem to indicate that both the solutions analyzed don't have a physical meaning for the minimal model. At
this point, in order to confirm the whole construction, we need to find at least one physical reflection matrix,
using the CDD-ambiguity mentioned.

\newpage

\resection{CDD-Ambiguity: the $E_{6}$ case}

We start considering the \lq\lq minimal\rq\rq \, reflection matrix (\ref{min1})-(\ref{min4}). The problem with the
poles $\theta=i\frac{3}{12}\pi$ in $K_{2}^{\mu}$ ($\mu=\beta_{1},\beta_{2},\gamma_{1},\gamma_{2}$) is that they
generate states characterized by a mixing of the Dynkin diagram colourations, because
\begin{equation}
S_{b2}\left(\theta+i\frac{3}{12}\pi\right)S_{b2}\left(\theta-i\frac{3}{12}\pi\right)=S_{b1}\left(\theta\right)S_{b6}\left(\theta\right).
\end{equation}
Their appearance is due to the fact that every minimal block $\{x\}$ evaluated at $\theta=i\frac{x}{h}\pi$ is a
negative quantity, hence it changes the residue sign as it enters the reflection amplitude expression at some
excited state. This doesn't happen in the Toda theory because another compensating negative sign arises from the
coupling-dependent factors. It is then clear that, adding a $S$-matrix factor to the initial reflection amplitude,
we will hardly overcome this problem, because we will get a more complicated pole structure and presumably we
will recover most of the previous boundary bound states.

The best strategy seems then to divide the reflection amplitude by an opportune $S$-matrix element. The Dynkin
colour which characterize the boundary bootstrap for the \lq\lq minimal\rq\rq \, solution is the one of particles
$2$, $3$ and $5$, and we will start with the simplest choice, i.e. the self-conjugate particle $2$. We then
define:
\begin{equation}\label{E6CDD}
\widetilde{K}_{a}^{0}(\theta)=S_{a2}^{-1}(\theta)K_{a}^{0}(\theta)
\end{equation}

The \lq\lq fusing angles\rq\rq \, are

\begin{center}
\begin{tabular}{|c|c|c|c|}\hline
 $ a\setminus\mu$  &  $ 0$  &  $\delta$ \\ \hline
  $1$ & $1^{\beta_{1}}$  & $(3)_{1}$   \\ \hline
  $2$ &   & $(2)_{1}\,6^{\delta}$   \\ \hline
  $3$ & $(3)_{2}$  &  $(1)_{2}(2_{3})_{2}(3)_{2}$  \\ \hline
  $4$ & $3^{\delta}$  &    \\ \hline
  $5$ & $(3)_{2}$  &  $(1)_{2}(2_{3})_{2}(3)_{2}$  \\ \hline
  $6$ & $1^{\beta_{2}}$  & $(3)_{1}$   \\ \hline
    \end{tabular}
\end{center}

\vspace{0.5cm}

where the brackets mean that the corresponding pole can be explained by Coleman-Thun diagrams, of the type
indicated by the external subscript. We have kept the same letters for the boundary states, because the relation
between the excited state reflection amplitudes and the ground state ones is the same as in the case of the
initial solution (\ref{min1})-(\ref{min4}).

The poles at $1^{\beta_{1,2}}$ in $K_{1,6}^{0}$ deserve a separate discussion. They would create again states on
which the reflection amplitude of particle 2 has a change of sign from negative to positive in the residue of the
pole at $i\frac{3}{12}\pi$, due to the appearance of the block $\{3\}$. We already know that Type 1 diagram
cannot be used, and there are not $b$ and $c$ such that $u_{bc}^{2}<\frac{\pi}{2}$. However, we know that if a
scattering amplitude $S_{bc}(\theta)$ has a pole at $\theta=i u_{bc}^{a}$, then the amplitude
$S_{\bar{b}c}(\theta)$ will have a corresponding \lq\lq crossed-channel\rq\rq \, pole at
$\theta=i(\pi-u_{bc}^{a})$. In this way, we could try to draw a \lq\lq crossed\rq\rq \, version of Type 2
diagram, represented by:

\vspace{5.5cm}

\begin{figure}[h]
\setlength{\unitlength}{0.0125in}
\begin{picture}(40,0)(60,470)

\thicklines \put(360,490){\line(0,1){140}}

\put(360,490){\line(2,1){15}} \put(360,500){\line(2,1){15}} \put(360,510){\line(2,1){15}}
\put(360,520){\line(2,1){15}} \put(360,530){\line(2,1){15}} \put(360,540){\line(2,1){15}}
\put(360,550){\line(2,1){15}} \put(360,560){\line(2,1){15}} \put(360,570){\line(2,1){15}}
\put(360,580){\line(2,1){15}} \put(360,590){\line(2,1){15}} \put(360,600){\line(2,1){15}}
\put(360,610){\line(2,1){15}} \put(360,620){\line(2,1){15}}

\put(360,560){\line(-2,1){30}} \put(300,590){\vector(2,-1){30}}

\put(360,560){\vector(-2,-1){40}} \put(320,540){\line(-2,-1){20}}

\put(300,530){\vector(0,1){40}} \put(300,570){\line(0,1){20}}

\put(300,590){\vector(-2,3){20}} \put(280,620){\line(-2,3){10}}

\put(300,530){\line(-2,-3){10}} \put(270,485){\vector(2,3){20}}

\put(290,560){$c$} \put(260,485){$a$} \put(260,630){$a$} \put(330,580){$\bar{b}$} \put(330,530){$\bar{b}$}

\end{picture}
 \caption{Crossed version of Type 2 diagram}
 \end{figure}
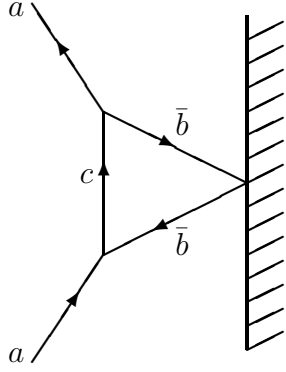

\vspace{0.5cm}

We will indicate the \lq\lq modified fusing angles\rq\rq \, of this diagram by a tilde; these are related to the
direct-channel ones in the following way:
\begin{eqnarray*}
\tilde{u}_{bc}^{a}&=&\pi-u_{bc}^{a}   \\
\tilde{u}_{ac}^{b}&=&u_{ac}^{b}   \\
\tilde{u}_{ab}^{c}&=&u_{ab}^{c}+(u_{bc}^{a}-\tilde{u}_{bc}^{a})=u_{ab}^{c}+(2u_{bc}^{a}-\pi)   \\
\tilde{\eta}_{a}&=&i(\tilde{u}_{bc}^{a}+\tilde{u}_{ab}^{c}-\pi)=i(u_{bc}^{a}+u_{ab}^{c}-\pi)   \\
\end{eqnarray*}

The boundary crossing equation (\ref{cross}) implies that the antiparticle reflection amplitude to be evaluated
at $\tilde{\eta}_{b}=\tilde{u}_{bc}^{a}$ is
\begin{equation}\label{crampl}
K_{\bar{b}}(\theta+i\pi)=\frac{S_{bb}(2\theta)}{K_{b}(\theta)}.
\end{equation}

\vspace{0.5cm}

This diagram can effectively explain the two poles in exam, with $(a,b,c)=(1,6,3)$ and $(a,b,c)=(6,1,5)$.

\newpage

The same idea can be applied to Type 3 diagram, when the version to be used is the one with \,
$\rm{Im}\left(\eta_{c}\right)>\frac{\pi}{2}$, which again requires $b$ and $c$ such that
$u_{bc}^{a}<\frac{\pi}{2}$. The \lq\lq modified fusing angles\rq\rq \, are defined as before, and the process is
represented by:

\vspace{5cm}

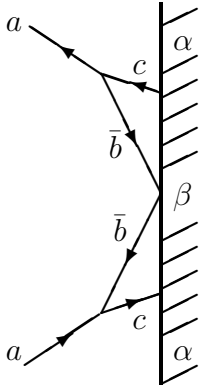
\begin{figure}[h]
\setlength{\unitlength}{0.0125in}
\begin{picture}(40,0)(60,470)
\thicklines \put(350,460){\line(0,1){160}}

\put(350,460){\line(2,1){15}} \put(350,480){\line(2,1){15}} \put(350,490){\line(2,1){15}}
\put(350,500){\line(2,1){15}} \put(350,510){\line(2,1){15}} \put(350,520){\line(2,1){15}}
\put(350,550){\line(2,1){15}} \put(350,560){\line(2,1){15}}

\put(350,570){\line(2,1){15}} \put(350,580){\line(2,1){15}} \put(350,590){\line(2,1){15}}
\put(350,610){\line(2,1){15}}

\put(350,540){\vector(-1,-2){15}} \put(335,510){\line(-1,-2){10}}

\put(325,590){\vector(1,-2){15}} \put(350,540){\line(-1,2){10}}

\put(350,582){\vector(-3,1){13}} \put(325,590){\line(3,-1){12}}

\put(325,490){\vector(3,1){17}}  \put(350,498){\line(-3,-1){12}}

\put(325,590){\vector(-3,2){18}}  \put(295,610){\line(3,-2){12}}

\put(293,468){\vector(3,2){20}}  \put(310,480){\line(3,2){14}}

\put(355,535){$\beta$} \put(355,470){$\alpha$} \put(355,600){$\alpha$}

\put(285,608){$a$} \put(285,470){$a$} \put(328,555){$\bar{b}$} \put(330,520){$\bar{b}$} \put(338,590){$c$}
\put(338,483){$c$}

\end{picture}
\caption{Crossed version of Type 3 diagram}
 \end{figure}

\vspace{0.5cm}

Also this crossed diagram, although of second order, has an immediate application to the two simple poles
mentioned, because now the reflection amplitudes can have zeroes in the physical strip, due to the $S_{a2}^{-1}$
term. If we choose $(a,b,c)=(1,5,4)$ (and $(a,b,c)=(6,3,4)$), $\alpha=0$ and $\beta=\delta$, we can describe the
poles at $\tilde{\eta}_{a}=i\frac{1}{12}\pi$, because particle 4 couples to the ground state at
$i\pi-\tilde{\eta}_{c}=i\frac{3}{12}\pi$ and
$\frac{S_{55}\left(2\theta\right)}{K_{5}^{\delta}\left(\theta\right)}=\frac{S_{33}\left(2\theta\right)}{K_{3}^{\delta}\left(\theta\right)}$
has a simple zero in $\tilde{\eta}_{b}=i\frac{2}{12}\pi$.

\vspace{0.5cm}

In this way, we have a bootstrap closing on the two states $0$ and $\delta$: the fact that we have skipped states
$\beta_{1,2}$ and $\gamma_{1,2}$ let us conclude that the same situation arises in the Toda theory and in the
minimal model.

\resection{Consequences of \lq\lq crossed diagrams\rq\rq}

It is now interesting to see what happens if we extend the use of \lq\lq crossed diagrams\rq\rq, fundamental in
the last discussion, also to the other sets of reflection amplitudes examined.

First of all we summarize the various possibilities given by the application of the direct-channel diagrams,
listing in the following table the number of states on which the boundary bootstrap closes (or presumably doesn't,
if we indicate \lq\lq$\infty$\rq\rq) in the various systems analyzed:

\begin{center}
\begin{tabular}{|c|c c|c c|c c|}\hline
&  $ E_{6}^{(1)}$  & $ E_{6}^{(2)}$ & $ E_{7}^{(1)}$ & $ E_{7}^{(2)}$ & $ E_{8}^{(1)}$ & $ E_{8}^{(2)}$ \\ \hline
Toda Field Theory  & 8 & 8 & 8 & 8 & 16 & 16 \\
Toda\,+\,Coleman-Thun  & 8 & 8 & 2 & 8 & 16 & 16 \\ \hline
Minimal Model (MM) & $\infty$ & $\infty$ & $\infty$ & $\infty$ & $\infty$ & $\infty$ \\
MM\,+\,Coleman-Thun & $\infty$ & $\infty$ & 2/8 & 8 & 16 & 16 \\ \hline

  \end{tabular}
\end{center}

\vspace{0.5cm}

The first column refers to the kind of theory: the voices \lq\lq Toda Field Theory\rq\rq \, and \lq\lq Minimal
Model \rq\rq \, discriminate between \lq\lq dressed\rq\rq \, and minimal scattering matrices, with a bootstrap
carried on all odd-order poles with positive residue, while with \lq\lq +\,Coleman-Thun\rq\rq \, we mean the
exclusion of boundary bound state creations if alternative diagrams can be drawn. The first row indicates the six
solutions examined (two for each algebra), with $(1)$ and $(2)$ referring respectively to the \lq\lq
minimal\rq\rq \, solution and to the one shifted by $i\pi$.

\vspace{0.5cm}

Obviously the new diagrams increase the number of explicable poles, most of all for light particles, but the
principal novelty is that their order can also be lowered, because in general the \lq\lq crossed\rq\rq \,
reflection amplitudes (\ref{crampl}) can have zeroes in the physical strip, even if this is impossible for the
\lq\lq direct-channel\rq\rq \, ones. However, this is true only if we are on excited states, because on the ground
state (\ref{crampl}) exactly corresponds to going from the \lq\lq minimal\rq\rq \, solution to the shifted one or
vice versa.

This implies that if a simple pole of $K_{b}^{0}$ cannot be explained by Coleman-Thun (normal or crossed) Type 1
or Type 2 diagrams, then we can conclude that it creates a boundary bound state, but this argument is not valid
on excited states. On these states, in fact, even if we don't find an opportune diagram to describe a certain pole
(simple or multiple), we cannot conclude that this pole corresponds to a boundary excitation, because we cannot
check all possible order diagrams with the opportune zeroes insertions.

\vspace{0.5cm}

We will now describe the combined effect of \lq\lq normal\rq\rq \, and \lq\lq crossed\rq\rq \, diagrams on the
various amplitudes considered.

\vspace{0.5cm}

Let's start with the $E_{6}$ \lq\lq minimal\rq\rq \, solution. In the Toda case we are able to reduce the number
of boundary states from 8 to 4, getting a bootstrap which closes on $0,\alpha,\beta_{1},\beta_{2}$. This doesn't
solve the minimal model problem, because it seems again that the poles at $\theta=i\frac{3}{12}\pi$ in
$K_{2}^{\beta_{1,2}}$ generate an infinite cascade of boundary states; we couldn't find opportune crossed
diagrams to describe this simple poles, but as we have explained we could not investigate all the possibilities.

As it regards the second solution, from the ground state we are sure to obtain states $\beta$ and $\gamma$,
skipping $\alpha_{1,2}$, but we cannot exclude the subsequent creation of $\delta_{1,2}$ and $\varepsilon$. Hence
for the Toda theory we can conclude that the bootstrap closes on a number of states between 3 and 6, while we
have not a definite answer for the minimal model.

\vspace{0.5cm}

For the other two algebras we already know that the use of standard Coleman-Thun diagrams gives the same boundary
states spectrum in the Toda theory and in the perturbed minimal model.

The bootstrap generated by the $E_{7}$ \lq\lq minimal\rq\rq \, solution remains unchanged on the two states $0$
and $\alpha$. With the second solution we certainly have the $\delta$ creation from the ground state (without
$\alpha$, $\beta$ and $\gamma$), but we cannot decide what happens with the following states $\varepsilon$,
$\sigma$ and $\tau$; the bootstrap will then close on a number of states between 2 and 5.

The $E_{8}$ case, finally, is very similar. The \lq\lq minimal\rq\rq \, solution generates states
$\alpha,\beta,\gamma,\delta,\varepsilon$ from the ground state, hence the bootstrap will close on a number of
states between 6 and all the 16. With the second solution, instead, we are sure to get states $\delta$ and
$\lambda$ avoiding $\alpha,\beta$ and $\gamma$, so that the uncertainty is between 3 and 13 states.

\vspace{0.5cm}

We summarize these results in the following table, with the same conventions as described before:

\begin{center}
\begin{tabular}{|c|c c c|c c|c c|}\hline
&  $ E_{6}^{(1)}$  & $ E_{6}^{(2)}$ & $ E_{6}^{(3)}$ & $ E_{7}^{(1)}$ & $ E_{7}^{(2)}$ & $ E_{8}^{(1)}$ & $
E_{8}^{(2)}$ \\ \hline
Toda  & 4 & 3-6 & 2 & 2 & 2-5 & 6-16 & 3-13 \\
MM & 4-$\infty$ & 3-$\infty$ & 2 & 2 & 2-5 & 6-16 & 3-13 \\ \hline
  \end{tabular}
\end{center}

\vspace{0.5cm}

Every entry of the form \lq\lq $n-m$\rq\rq \, means that we cannot decide on how many states the bootstrap
closes, but this number should lie between $n$ and $m$. The additional solution indicated by $E_{6}^{(3)}$ is the
one with ground state amplitudes (\ref{E6CDD}).

\vspace{0.5cm}

\resection{Discussion}

We have performed a detailed analysis of the boundary states structures arising from the reflection amplitudes
found by Fring and Koberle, showing how generalized Coleman-Thun mechanisms can have interesting consequences
compared with a blind iteration of the bootstrap on all odd-order poles with positive residue. However, these
on-shell methods are not sufficient to outline a clear and definitive picture of the phenomenon. We are in fact
left with various kinds of problems.

The first one is to understand if the possibility of drawing generalized Coleman-Thun diagrams really excludes the
creation of a boundary bound state. This seems reasonable in cases where the bootstrap closes only under this
assumption, as for perturbed minimal models, but we have seen that, for the \lq\lq minimal\rq\rq \, solution in
the $E_{7}$ Toda theory, the alternative is between two closing bootstraps, one on eight states, and the other on
two. Another eventuality is that the same ground state reflection amplitudes correspond to distinct boundary
conditions, whose different physical properties become manifest in the interpretation of certain poles. This
phenomenon was recognized in \cite{tateo} in the case of the scaling Lee-Yang model, knowing independently the
different spectra from a boundary generalization of the truncated conformal space approach.

The direct strategy to face this problem would be to calculate the residues of the various diagrams and compare
them with the actual residue of the corresponding pole in the reflection amplitude, but it is not known how to
treat this perturbative calculations in the presence of a boundary.

Another delicate point is the use of crossed diagrams, which are so important in the $E_{6}$ case; again it would
be necessary to calculate their contribute to the residues. Furthermore, the possibility of inserting in these
diagrams crossed reflection amplitudes with zeroes in the physical strip makes it very difficult to conclude
something about many poles, and alternative methods are essential to check the existence of the related boundary
states.

Finally, in this context we have no way to understand which is the boundary condition related to a certain
reflection matrix, and the best we can do is just to notice whether an eventual symmetry of the systems is
preserved or not by the corresponding excited boundary states structure. This is a particularly delicate problem,
especially in the light of the difference between the bound states spectra displayed by the minimal and the Toda
reflection amplitudes. It could be, in fact, that the two basic solutions analyzed in the $E_{6}$ case are
related to boundary conditions that in the UV limit correspond to conformal boundary conditions present in the CFT
associated to the Toda Lagrangian but not in the tricritical 3-state Potts model.

\vspace{0.5cm}

To solve these problems, we intend to proceed in the future work performing indirect checks on certain properties
of the theory, using TBA equations (\cite{TBA1},\cite{TBA2}) and analyzing one-point function behaviours
(\cite{FF}). The basic idea is to perform the UV limit in order to calculate the $g$-function (\ref{g}) related
to a certain set of reflection amplitudes. The various possible choices of excited boundary states should produce
different values of $g$, which in the case of minimal models can be compared to the ones given by the primary
operators present in the CFT using eq.(\ref{valg}). In this way it should be possible to identify the physical
meaning of the reflection matrices, associating them to specific boundary conditions, and to deduce which of the
initial boundary bound states are really involved in the bootstrap.

\vspace{1.5cm}

\textbf{Acknowledgements}. I'm grateful to G. Delfino, D. Fioravanti and R. Tateo for very useful discussions and
suggestions, and to G. Mussardo for having introduced me into the field and for his continuous interest and
contribution to the development of this work. I also thank S.I.S.S.A. for a pre-graduate grant and for
hospitality.

\end{document}